\documentclass[a4paper,11pt]{article}
\pdfoutput=1 % if your are submitting a pdflatex (i.e. if you have
\usepackage{jcappub} % for details on the use of the package, please
\usepackage{epsfig}
\usepackage{url}
\usepackage{hyperref}
\hypersetup{citecolor=black,linkcolor=black}

\usepackage[normalem]{ulem} 

\usepackage{latexsym}
\usepackage{epsfig}
\usepackage{amsmath}
\usepackage{amssymb}
\usepackage{bm}
\usepackage{graphicx}
\usepackage{verbatim}
\usepackage{enumerate,mdwlist}%lists
\usepackage[titletoc]{appendix}
\usepackage{amsfonts}
\usepackage{tikz} %diagrams

\newcommand{\lp}{\left(}
\newcommand{\rp}{\right)}
\newcommand{\lb}{\left[}
\newcommand{\rb}{\right]}

\newcommand{\ba}{\begin{align}}
\newcommand{\ea}{\end{align}}
\newcommand{\be}{\begin{equation}}
\newcommand{\ee}{\end{equation}}

\newcommand{\PS}{\mathcal{P}}%Power Spectrum

\newcommand{\bpm}{\begin{pmatrix}}%matrices
\newcommand{\epm}{\end{pmatrix}}%matrices

\newcommand{\eps}{\epsilon_1}
\newcommand{\epss}{\epsilon_2}
\newcommand{\epsss}{\epsilon_3}
\newcommand{\epsn}{\epsilon_n}

\newcommand{\pip}{\bar{\pi}_\phi}
\newcommand{\bphi}{\bar{\phi}}
\newcommand{\dpi}{\delta\pi}
\newcommand{\dphi}{\delta\phi}

\newcommand{\kv}{\vec{k}}
\newcommand{\xv}{\vec{x}}
\newcommand{\bPhi}{\mathbf{\Phi}}

\newcommand{\bu}{\mathbf{u}}
\newcommand{\bx}{\mathbf{x}}

\newcommand{\bv}{\begin{pmatrix}}
\newcommand{\ev}{\end{pmatrix}}

\newcommand{\order}[1]{\mathcal{O}\lp#1\rp}

\newcommand{\mpp}[2]{\langle\dphi^{#1}\dpi^{#2}\rangle} %phi\pi
\newcommand{\mppp}[3]{\langle\dphi^{#1}\dpi^{#2}#3\rangle} % \phi\pi something
\newcommand{\mphi}[1]{\langle\dphi^{#1}\rangle}
\newcommand{\mpphi}[2]{\langle\dphi^{#1}#2\rangle}
\newcommand{\mpi}[1]{\langle\dpi^{#1}\rangle}

\newcommand{\mph}{\langle\dphi\rangle}
\newcommand{\mpii}{\langle\dpi\rangle}
\newcommand{\mphp}{\langle\dphi\dpi\rangle}
\newcommand{\mphipin}[1]{\langle\dphi\dpi^{#1}\rangle}
\newcommand{\mphinpi}[1]{\langle\dphi^{#1}\dpi\rangle}

\definecolor{grey}{rgb}{0.4,0.4,0.4}
\definecolor{dullmagenta}{rgb}{0.4,0,0.4}
\definecolor{darkblue}{rgb}{0,0,0.4}
\definecolor{midblue}{rgb}{0,0,0.5}
\definecolor{midred}{rgb}{0.5,0,0}
\definecolor{orange}{rgb}{1,0.5,0}
\definecolor{lightbrown}{rgb}{0.75,0.5,0.25}
\definecolor{tan}{cmyk}{0.14,0.42,0.56,0}
\definecolor{djunglegreen}{cmyk}{0.99,0,0.52,0}
\definecolor{lightgreen}{rgb}{0,1,0}
\definecolor{olivegreen}{cmyk}{0.64,0,0.95,0.40}
\definecolor{midgreen}{rgb}{0.0,0.675,0.0}
\definecolor{darkgreen}{rgb}{0,0.5,0}

\usepackage[all]{xy} %array diagrams
\usepackage{amsfonts}

\author[a,b]{Jose Mar\'ia Ezquiaga}
\affiliation[a]{Instituto de F\'isica Te\'orica UAM-CSIC, Universidad Aut\'onoma de Mardid, Cantoblanco, Madrid, 28049 Spain}
\affiliation[b]{Theoretical Physics Department, CERN, 1211 Geneva 23, Switzerland}
\emailAdd{jose.ezquiaga@uam.es}
\author[a,b]{and Juan Garc\'ia-Bellido}
\emailAdd{juan.garciabellido@uam.es}

\title{Quantum diffusion beyond slow-roll: implications for primordial black-hole production}

\abstract{Primordial black-holes (PBH) can be produced in single-field models of inflation with a quasi-inflection point in the potential. In these models, a large production of PBHs requires a deviation from the slow-roll (SR) trajectory. In turn, this SR violation can produce an exponential growth of quantum fluctuations. We study the back-reaction of these quantum modes on the inflationary dynamics using stochastic inflation in the Hamilton-Jacobi formalism. We develop a methodology to solve quantum diffusion beyond SR in terms of the statistical moments of the probability distribution. We apply these techniques to a toy model potential with a quasi-inflection point. We find that there is an enhancement of the power spectrum due to the dominance of the stochastic noise in the phase beyond SR. Moreover, non-Gaussian corrections become as well relevant with a large positive kurtosis. Altogether, this produces a significant boost of PBH production. We discuss how our results extend to other single-field models with similar dynamics. We conclude that the abundance of PBHs in this class of models should be revisited including quantum diffusion.
}
\date{\today}

\keywords{physics of the early universe, inflation, primordial black holes}

\begin{document} 

\maketitle
\flushbottom

%--------
%SEC 1: INTRODUCTION
%--------
\section{Introduction}

The tale of modern cosmology is rather elegant. Large scale structures of the universe are seeded by quantum fluctuations generated during the first instants and stretched to cosmological scales by the exponential expansion of inflation. Yet, quantum fluctuations could also have a decisive role in the small scales of the universe. This would be the case if there was an enhancement of the power spectrum at small scales, triggering the collapse of large fluctuations into primordial black-holes (PBH) \cite{Carr:1974nx}. PBHs are interesting objects because they leave imprints throughout the history of the universe by their energy injections, dynamical effects and gravitational waves, topics reviewed recently in \cite{Carr:2016drx,Garcia-Bellido:2017fdg,Sasaki:2018dmp}. They could also help in solving the origin of the black-holes detected by LIGO \cite{Bird:2016dcv,Clesse:2016vqa,Sasaki:2016jop} or the seeds for supermassive black-holes \cite{Clesse:2015wea,Carr:2018rid}. If they are abundant enough, they could comprise a large fraction of the dark matter \cite{Clesse:2017bsw}. 

The fraction of PBH formed in the early universe $\beta_f$ is determined by the probability that a given primordial curvature fluctuation $\zeta$ is above a certain threshold
\be
\beta_f(M)=\int_{\zeta_c}^{\infty}P(M;\zeta)d\zeta.
\ee
Therefore, the abundance of PBHs is both sensitive to the probability density function (PDF) $P(M;\zeta)$ and the value of the threshold $\zeta_c$. The PDF is governed by the physics in the early universe while the threshold is subject to the conditions at the time of formation. The analysis of the formation of PBHs and the appropriate threshold condition has been an active line of research in the past years \cite{Kopp:2010sh,Harada:2013epa,Young:2014ana,Germani:2018jgr,Yoo:2018kvb}. Since the purpose of this work is to focus on the computation of the PDF from the inflationary dynamics, we will follow the above prescription for the formation of PBHs and fix $\zeta_c=0.5$. Using another prescription might change the numbers for $\beta_f$ but would not change the results qualitatively. We are interested in comparing the production of PBHs with and without quantum diffusion.

Large curvature fluctuations can be produced by very different means in the early universe. We will be interested in producing them during inflation. Within inflation, we will focus on single-field models. There, one can produce large fluctuations if there is a second plateau in the potential \cite{Garcia-Bellido:2017mdw}. Examples of single-field models producing PBHs are critical Higgs inflation \cite{Ezquiaga:2017fvi}, double inflation \cite{Kannike:2017bxn}, radiative plateau inflation \cite{Ballesteros:2017fsr} or some string realizations \cite{Cicoli:2018asa,Ozsoy:2018flq,Dalianis:2018frf}. In this class of models, in order to have sufficiently large fluctuations, the inflationary dynamics has to deviate from slow-roll (SR) \cite{Kannike:2017bxn,Germani:2017bcs,Motohashi:2017kbs}. As we will review, this period beyond SR leads to an exponential growth of the quantum modes \cite{Leach:2000yw,Leach:2001zf}.

Since there will be a regime with large quantum fluctuations, it is natural to explore the back-reaction of these quantum fluctuations on the classical inflationary dynamics and if they affect the production of PBHs. This will be the main objective of this work. For this goal, stochastic inflation \cite{Starobinsky:1986fx} is the appropriate tool since it tracks stochastic effects of the short wavelength modes on the long wavelength ones. Because in this class of models there are deviations from SR, we have to consider the stochastic evolution of both the coarse-grained field $\bphi$ and its canonical momentum $\pip$ with their respective noises $\xi_\phi$ and $\xi_\pi$ \cite{Tolley:2008na,Grain:2017dqa}. Then, we will obtain the PDF of the stochastic fluctuations from their correlation functions solving a system of first-order differential equations.
An alternative route for computing the statistical moments would be to consider the $\delta N$ formalism in which $n$-point correlation functions have been computed for SR inflation \cite{Vennin:2015hra} and for multi-field inflation \cite{Assadullahi:2016gkk,Vennin:2016wnk}.
 
One should note that higher order correlators can be very significant because the fraction of PBHs formed is very sensitive to the tails of the PDF. Therefore, it is important to compute both Gaussian and non-Gaussian (NG) contributions. Several studies have discussed extensively the importance of considering NG corrections to PBH formation \cite{Saito:2008em,Byrnes:2012yx,Young:2013oia,Young:2015cyn,Tada:2015noa,Young:2015kda,Franciolini:2018vbk}. In our case, they will be relevant too. Fortunately, NG contributions to the abundance of PBHs $\beta_f$ can be computed analytically from the correlation functions \cite{Matarrese:1986et}.
Subsequently, once $\beta_f$ is in hand, the effects of quantum diffusion in the production of PBHs can be discussed. Such effects have been considered previously in the context of hybrid inflation \cite{GarciaBellido:1996qt,Clesse:2015wea} and for single-field SR inflation \cite{Bullock:1996at,Ivanov:1997ia,Yokoyama:1998pt,Pattison:2017mbe}. In both cases, quantum diffusion was found to be a key player. Recently, the necessity of considering quantum diffusion for models with SR violation has been pointed out in \cite{Biagetti:2018pjj}. 

Here, we place a full analysis of the quantum diffusion beyond SR, computing the noise and the non-Gaussian corrections. We find  that quantum diffusion can significantly alter the classical prediction. To arrive at this result, we first review in Sec. \ref{sec:Inflation} how an enhancement of the power spectrum occurs in single-field models with a quasi-inflection point. Then, we present stochastic inflation in the Hamilton-Jacobi formalism in Sec. \ref{sec:Stochastic}. Subsequently, in Sec. \ref{sec:Correlation functions}, we develop a methodology to solve the problem by calculating the statistical moments of the PDF. We present our results for the abundance of PBHs in Sec. \ref{sec:PBH}, discussing how our findings extend to other models. We conclude in Sec. \ref{sec:Conclusions}. 

%--------
%SEC 2: INTRODUCTION
%--------
\section{Inflation with a quasi-inflection point}
\label{sec:Inflation}

We are interested in studying inflationary scenarios with an quasi-inflection point because they naturally give rise to a peak in the power spectrum \cite{Garcia-Bellido:2017mdw}. In order to understand the dynamics of the inflaton in these scenarios it is convenient to parametrize the evolution in terms of the Hubble flow and its derivatives with respect to the number of $e$-folds $dN=Hdt$. For that purpose, we use the following Hubble-flow parameters
\be
\label{eq:epsn}
\epsn=\frac{d\ln \vert\epsilon_{n-1}\vert}{dN}\,,
\ee
where the initial parameter in the series corresponds to the inverse Hubble parameter $\epsilon_0=1/H$. Then, the first parameter is $\eps=-H(N)'/H$. Within this language, slow-roll inflation is defined by $\vert\epsn\vert\ll1$.

The fact that there is a quasi-inflection point $\phi_c$ in the potential is given by the condition $V_{,\phi}(\phi_c)\approx V_{,\phi\phi}(\phi_c)\approx0$, where $_{,\phi}$ indicates a partial derivative with respect to the inflaton $\phi$. For simplicity we will describe inflation with a toy model potential \cite{Garcia-Bellido:2017mdw}
\be
\label{eq:potential}
V(\phi)=\frac{1}{12}\frac{6m^2\phi^2-4\alpha\phi^3+3\lambda\phi^4}{\lp1+\xi\phi^2\rp^2}\,,
\ee
and choose the parameters to fulfill the quasi-inflection point condition accordingly\footnote{In particular, we use in the plots $\lambda=1$, $\xi=2.3$, $\alpha=6\lambda\,\phi_c/(3+\xi^2\phi_c^4)-4.3\cdot10^{-5}$ and $m^2=\lambda\phi_c^2(3+\xi\phi_c^2)/(3+\xi^2\phi_c^4)$, with the field starting at $\phi_{65}=6.5\phi_c$.}. Note that we are interested in a quasi-inflection point rather than a true inflection point, $V_{,\phi}=V_{,\phi\phi}=0$, since we want the inflaton to continue rolling ending inflation at the minimum of the potential avoiding getting trapped in the self-reproduction regime \cite{Allahverdi:2006iq}. A graphical representation of this potential can be found in the left panel of Fig. \ref{fig:Potential}. The $\lp1+\xi\phi^2\rp^2$ function in the denominator is introduced to flatten the potential at large field values, improving the agreement with CMB observations. Within the context of the $\epsn$ parameters, the spectral index $n_s$ and the tensor-to-scalar ratio $r$ are given respectively by $n_s=1-2\eps-\epss$, $r=16\eps$.
Now, before introducing the effects of quantum diffusion, let us first review the classical dynamics of this toy model and how an enhancement of the power spectrum is produced.

%Classical
\subsection{Classical dynamics}

The classical dynamics of the inflaton fields is determined by its evolution equation
\be
\frac{d^2\phi}{dN^2}+(3-\eps)\frac{d\phi}{dN}+\frac{(3-\eps)}{\kappa^2}\lp\ln V\rp_{,\phi}=0\,.
\ee
At large field values the inflaton slow-rolls the potential giving rise to the CMB fluctuations. Before crossing the quasi-inflection point, the inflaton acquires some inertia that is rapidly lost when the inflaton gets closer and starts feeling the flatness of the potential around the quasi-inflection point. This makes the inflaton to spend many $e$-folds of its evolution around this quasi-inflection point. During this time, also known as ultra slow-roll \cite{Namjoo:2012aa,Martin:2012pe,Chen:2013aj}, fluctuations would be largely amplified and could lead to the production of PBH. However, for suitable potential, it retains enough inertia to cross this point and end inflation oscillating around the minimum of the potential.

\begin{figure}[!t]
\centering
\includegraphics[width = 0.49\textwidth]{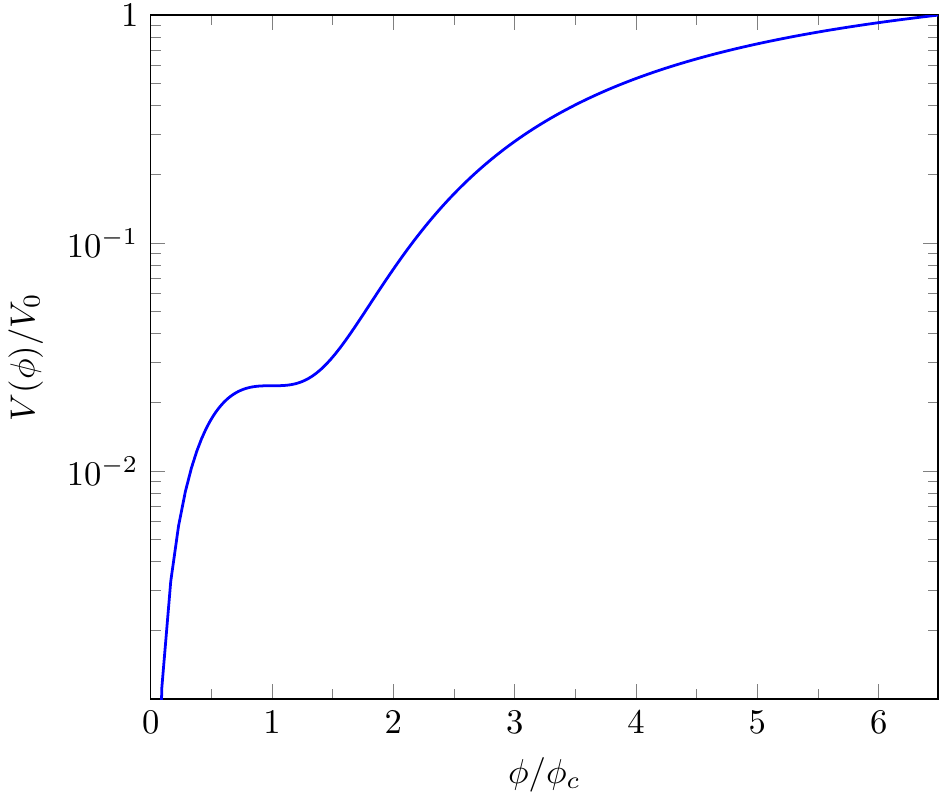}
\includegraphics[width =0.47 \textwidth]{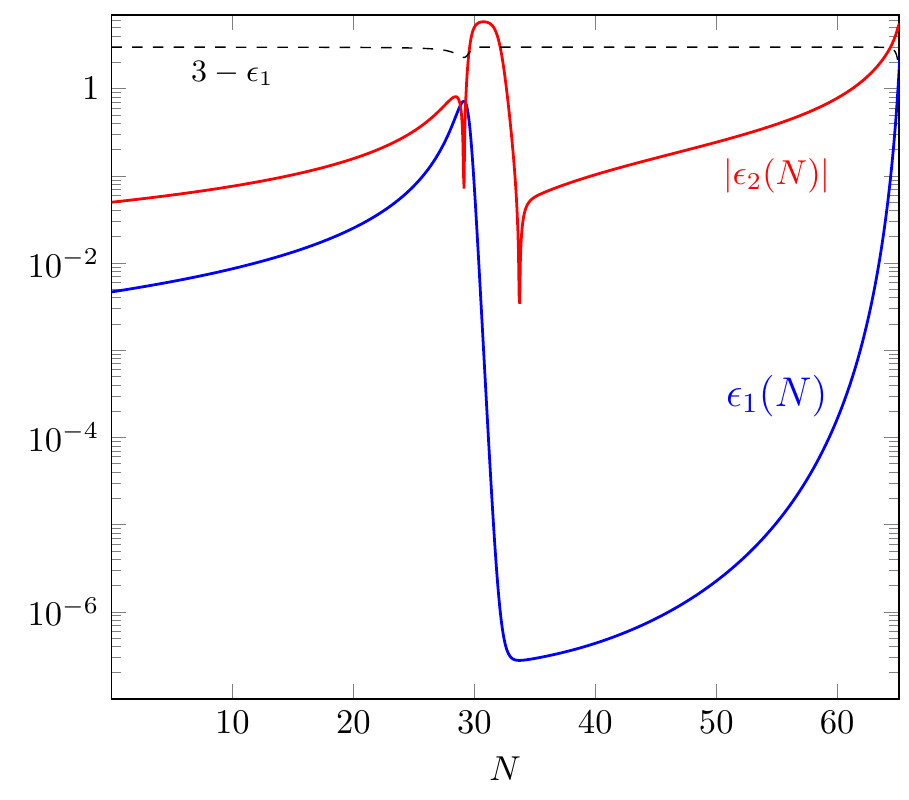}
\caption{On the left, inflationary potential with a quasi-inflection point at $\phi_c$ for the toy model (\ref{eq:potential}). On the right, evolution of the first slow-roll parameters $\eps$ and $\epss$, cf. (\ref{eq:epsn}). The dashed line corresponds to $3-\eps$. In the region where $\vert\epss\vert$ is above this line, there will be an enhancement of the power spectrum after horizon crossing.}
\label{fig:Potential}
\end{figure}

This characteristic evolution has two implications. Firstly, since the inflaton slows down around the quasi-inflection point, the first Hubble-flow parameter $\eps$ will become very small. This will introduce an enhancement of the power spectrum since $\PS_\zeta\sim1/\eps$. Secondly, the fact that the inflaton changes its velocity very rapidly as it gets to the quasi-inflection point will bring its evolution outside of slow-roll. This can be seen by the fact that $\vert\epss\vert$ becomes large. In turn, this can produce a further enhancement of the power spectrum by exciting growing modes outside the horizon. A characterization of this specific behavior of $\eps(N)$ and $\epss(N)$ in these scenarios is plotted in the right panel of Fig. \ref{fig:Potential}.

% FIG. 1
%

%Enhanced Power spectrum
\subsection{Enhancement of the power spectrum}
\label{sec:EnhancementPS}

In order to obtain the primordial power spectrum of inflation we need to solve the evolution of the inflaton and metric quantum fluctuations. For that purpose, we use the gauge invariant curvature fluctuation
\be
\zeta=\frac{u}{z}=\frac{a\delta\phi}{z}\,,
\ee
where $u$ is the Mukhanov-Sasaki variable \cite{Mukhanov:1985rz,Sasaki:1986hm} and we define the classical field $z=a\, d\phi/dN$. Note that we are absorbing the scalar metric perturbation $\Phi$ in the inflaton fluctuations $\delta\phi$ by adding a correction to the time dependent, effective mass term. This is equivalent to work in the uniform-curvature gauge.

In Fourier space, the evolution of the curvature perturbations $\zeta_k=u_k/z$ is described by\footnote{This can be easily derived from the usual Mukhanov-Sasaki equation $u_k''(\eta)+(k^2-z''/z)u_k=0$ by changing from conformal time to the number of $e$-folds, $dN=\mathcal{H}d\eta$. Also, we have used that $d\ln z/dN=1+\epss/2$.},
\be
\label{eq:MSpert}
\frac{d^2\zeta_k}{dN^2}+\lp3-\eps+\epss\rp\frac{d\zeta_k}{dN}+\lp\frac{k}{aH}\rp^2\zeta_k=0\,.
\ee
This equation has two well-defined regimes delimited by the comoving horizon size $d_{_\text{H}}=1/aH$. At sub-horizon scales ($k\gg aH$), the friction term is irrelevant and the equation describes a free field in Minkowski space, which can be normalized as a Bunch-Davis vacuum. At super-horizon scales ($k\ll aH$), the asymptotic solution is
\be
\left.\frac{d\zeta_k}{dN}\right\vert_{k\ll aH}=C_2e^{-\int(3-\eps+\epss)dN}=\tilde{C}_2e^{-3N+\ln H-\ln\eps}\,,
\ee
implying that there are two modes: one constant and another evolving
\be
\zeta_{k\ll aH}=C_1+\tilde{C}_2\int e^{-3N+\ln H-\ln\eps}dN\,.
\ee
Therefore, depending on the sign of $3-\eps+\epss$ the second mode will be exponentially decaying or growing. The power spectrum associated to the curvature perturbations is then 
\be
\label{eq:PMS}
\PS^{_\text{MS}}_\zeta=\left.\frac{k^3}{2\pi^2}\vert\zeta_k\vert^2\right\vert_{k\ll aH}\,,
\ee
which is obtained solving numerically the mode equation (\ref{eq:MSpert}). Interestingly, the mode equation could be rewritten as a system of two first-order equations for the adiabatic and entropy perturbations \cite{Leach:2000yw}, showing that the only source of growth of the curvature after horizon crossing are the entropy perturbations.

Usually this evolving mode is exponentially suppressed because $\eps$ and $\epss$ are small. Then the curvature perturbation becomes constant very rapidly after horizon crossing and its power spectrum can be evaluated at this time $k=aH$. The power spectrum itself can be computed solving the equation for the curvature perturbation with the proper normalization. Doing that one arrives at the standard result 
\be
\label{eq:PSR}
\PS^{_\text{SR}}_\zeta\simeq \frac{\kappa^2}{8\pi^2}\frac{V(\phi)}{\eps(3-\eps)}\,,
\ee
where the $\simeq$ indicates that this is not an exact result beyond slow-roll. From this formula one can easily elucidate a method to amplify the fluctuations, i.e. slow down the inflaton to reach a small value of $\eps$ since the power spectrum would scale as $\PS_\zeta\sim1/\eps$. This enhancement due to $1/\eps$ was used to form PBH in \cite{Garcia-Bellido:2017mdw,Ezquiaga:2017fvi}. Note that in order to have a significant peak one needs to go beyond slow-roll \cite{Kannike:2017bxn,Germani:2017bcs,Motohashi:2017kbs} and the power spectrum (\ref{eq:PSR}) might not be a good approximation \cite{Motohashi:2017kbs,Ballesteros:2017fsr}.

\begin{figure}[!t]
\centering
\includegraphics[width = 0.477\textwidth]{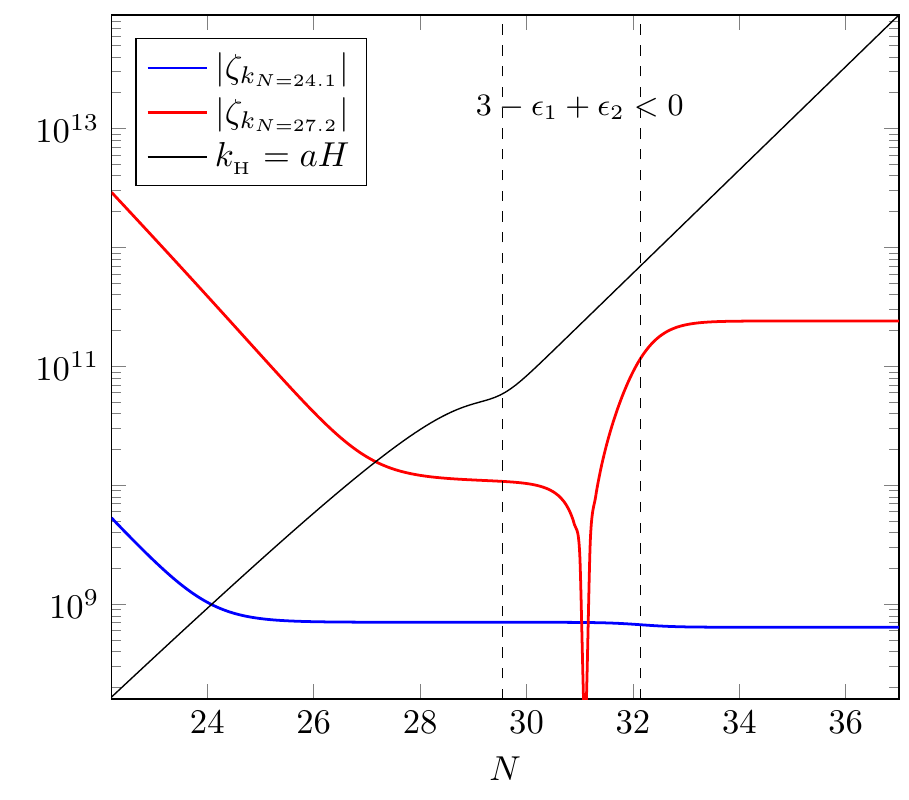}
\includegraphics[width = 0.48\textwidth]{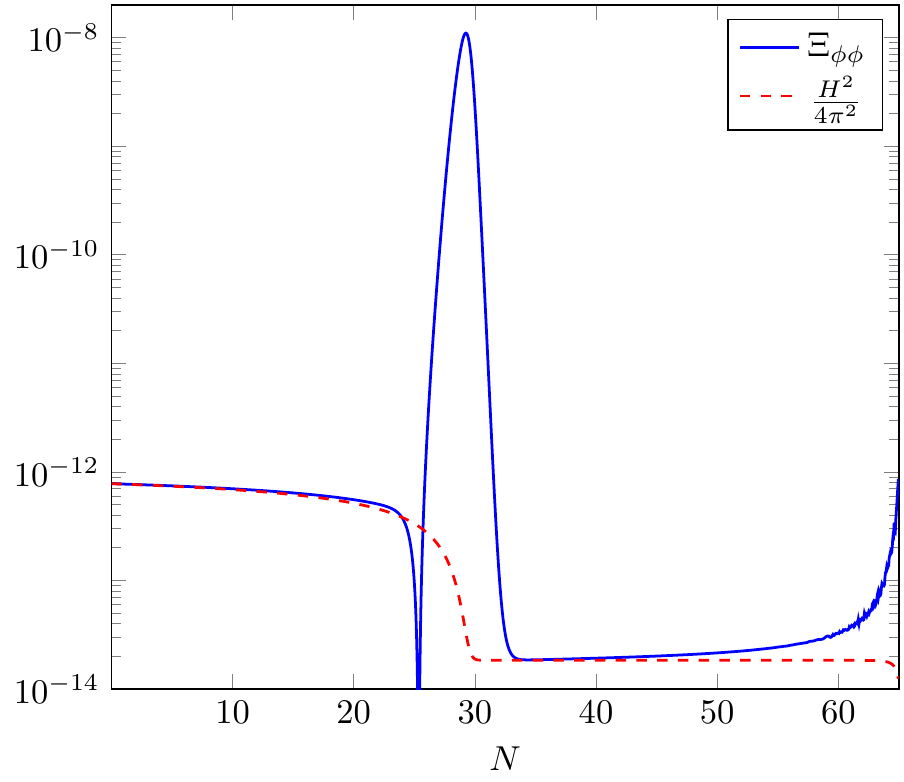}
\caption{On the left, evolution of two curvature perturbations $\zeta_k$ that exit the horizon $k=aH$ at different times. There is an enhancement of the modes after horizon crossing in the region between dashed lines where $3-\eps+\epss<0$ is satisfied. This produces an enhancement of the stochastic noise $\Xi_{\phi\phi}$ in this region with respect to the slow-roll result $H^2/4\pi^2$ as shown in the right panel.}
\label{fig:Modes}
\end{figure}

However, if we are beyond slow-roll, the friction term could change sign and the non-adiabatic mode would grow exponentially \cite{Leach:2000yw,Leach:2001zf}. This will happen whenever $\epss<-3+\eps$ and the decaying mode will become a growing mode. Thus, the curvature perturbation $\zeta$ will grow after horizon crossing. This growth on super-horizon scales is therefore another way to enhance the power spectrum.\footnote{Nevertheless, one should note that this enhancement on super-horizon scales does not occur for the tensor perturbations \cite{Leach:2000yw} because their friction does not flip sign, since they follow the evolution equation $\frac{d^2v_k}{dN^2}+(1-\eps)\frac{dv_k}{dN}+\lb\lp\frac{k}{aH}\rp^2-(2-\eps)\rb v_k=0$ where $v_k=a\,h_k$.} This enhancement after horizon crossing was already considered as a source of PBH in \cite{Leach:2000yw}. More recently, this enhancement has been shown to be important for different single-field inflationary models of PBH production \cite{Saito:2008em,Ballesteros:2017fsr,Cicoli:2018asa,Ozsoy:2018flq}. For our toy model (\ref{eq:potential}), there will also be such growth after horizon crossing. This can be seen in the left panel of Fig. \ref{fig:Modes}, where we have presented the evolution of two different $\zeta_k$ exiting the horizon at different times. Modes that exit the horizon near the region or in the region where $3-\eps+\epss<0$ will suffer from this rapid growth. On the contrary, modes that exit the horizon well before or after this region will soon become constant after horizon crossing.

The fact that there is a strong enhancement of quantum fluctuations during the period beyond slow-roll will have important consequences for quantum diffusion. This is because then the quantum diffusion can dominate over the classical drift. In the next sections we will study the back-reaction of these large quantum fluctuations on the classical inflationary dynamics using the formalism of stochastic inflation beyond slow-roll. We will see that quantum diffusion can further enhance the power spectrum. Moreover, non-Gaussian correction will also enter in the game.  

% FIG. 2
%

%----------
%SEC 3: STOCHASTIC INFLATION BEYOND SR
%----------
\section{Stochastic inflation beyond slow-roll}
\label{sec:Stochastic}

Once we have studied the classical dynamics of inflation with a quasi-inflection point, it is natural to ask if there will be quantum corrections modifying the inflationary observables. For that purpose, stochastic inflation is the appropriate tool since it tracks the effects of the short-wavelength modes exiting the horizon on the long-wavelength ones by coarse-graining the field \cite{Starobinsky:1986fx,Linde:1993xx}. Thus, this formalism captures stochastically quantum correction to the classical evolution, which is exactly our goal. 

Since we are studying a system beyond slow-roll, we have to solve for both the field and its canonical momentum. Therefore, it is appropriate to work in phase space using the Hamilton-Jacobi formalism, see for instance \cite{Salopek:1990jq,GarciaBellido:1994vz} for applications to inflation. The analysis of stochastic corrections to the phase space dynamics has been addressed previously in Ref. \cite{Tolley:2008na,Grain:2017dqa}.
In this first introduction to stochastic inflation we will follow closely the developments of Ref. \cite{Grain:2017dqa}. We use the number of $e$-folds as time variable\footnote{See a detailed discussion in Ref. \cite{Vennin:2015hra} of why this is the appropriate variable.}, $dN=Hdt$, and define the conjugate momenta $\pi_\phi\equiv d\phi/dN$. We then split the field in long and short wavelengths as
\begin{align}
&\phi=\bphi+\phi_s \label{eq:splitphi} \\ 
&\pi_\phi=\pip+(\pip)_s\,, \label{eq:splitpi}
\end{align}
where quantities with an over bar represent the coarse-grained field and the subindex $s$ indicates the short-wavelength modes. The evolution of the long wavelengths can be obtained by solving the Langevin equations
\begin{align}
\label{eq:Langevin}
\frac{d\bphi}{dN}&=\pip+\xi_{\phi}\,, \\
\frac{d\pip}{dN}&=-(3-\epsilon)\lp\pip+\kappa^{-2}(\ln V)_{,\phi}\rp+\xi_\pi\,,
\end{align}
where $\xi_\phi$ and $\xi_\pi$ are the noise associated to $\bphi$ and $\pip$ respectively. The noise terms encapsulate all the physics associated to the short-wavelength modes. They can be computed by integrating the quantum modes in Fourier space with a time dependent cut-off tracking the horizon size. We leave the detailed derivation of the noise in Appendix \ref{app:noise}. The important result is that the 2-point correlations of the noise can be associated to the power spectrum of the fluctuations \cite{Grain:2017dqa}
\be \label{eq:noisecorrelation}
\Xi_{AB}(N,N')\equiv\langle0\vert\xi_A(N)\xi_B(N')\vert0\rangle=\frac{d\ln k_\sigma}{dN}\PS_{AB}(k_\sigma,N)\cdot\delta(N-N')\,,
\ee
where $k_\sigma=\sigma aH$ is the cutoff (with $\sigma\ll1$) and $A,B$ could be either $\delta \phi$ or $\delta\pi$. We are defining the power spectrum as
\be
\PS_{AB}(k_\sigma,N)=\frac{k_\sigma^3(N)}{2\pi^2}A_{k_\sigma}(N)B^\star_{k_\sigma}(N)\,.
\ee
For instance, the noise associated to the coarse-grained field $\bphi$ is characterized by
\be
\Xi_{\phi\phi}(N)=(1-\eps(N))\PS_{\delta\phi}(N)\,,
\ee
since $d\ln k_\sigma/dN=1-\epsilon_1$. One should note that in general there will be also a contribution to the noise in the momentum, $\Xi_{\pi\pi}\neq0$, and a correlation between them, $\Xi_{\phi\pi}\ne0$. It is only in the slow-roll approximation that $\Xi_{\pi\pi}\approx\Xi_{\phi\pi}\approx0$. The fact that the noise is defined through its 2-point function (\ref{eq:noisecorrelation}) is a consequence of working at linear order in perturbation theory for the short-wave modes. Moreover, the noise is uncorrelated at different times, $\Xi_{AB}(N,N')\propto\delta(N-N')$, i.e. it is white/Markovian noise. This is a result of choosing a Heaviside window function as the momentum cutoff \cite{Tolley:2008na,Grain:2017dqa} (see details in App.~\ref{app:noise}). Alternative choices still give a correlation which is exponentially suppressed at times $\Delta N=N-N'\geq1$ \cite{Winitzki:1999ve}.

The power spectrum of the fluctuations can be computed by solving the linearized equations for the short-wavelength modes (in phase space)
\begin{align}
\frac{d\delta\phi_k}{dN}&=\delta\pi_k\,, \label{eq:modephi} \\
\frac{d\delta\pi_k}{dN}&=-(3-\eps)\delta\pi_k-\lb\lp\frac{k}{aH}\rp^2+\frac{3-\eps}{\kappa^2}\frac{V_{,\phi\phi}}{V}-2\eps(3-\eps+\epss)\rb\delta\phi_k\,. \label{eq:modepi}
\end{align}
Note that we have included in the effective mass term a piece $-2\eps(3-\eps+\epss)$ that accounts for the effect of the metric fluctuations and, thus, the curvature perturbation obeys $\zeta_k=\kappa\delta\phi_k/\sqrt{2\eps}$. Then, the spectrum derived from the above evolution equations is equivalent to the Mukhanov-Sasaki formalism\footnote{In fact, it is easy to find the equivalence between both approaches if $V_{,\phi\phi}$ is expressed in terms of slow-roll parameters $\eps$ using that $V=(3-\eps)H^2/\kappa^2$ and $d\phi/dN=\sqrt{2\eps}/\kappa$.}. The noise associated to $\bphi$ in our inflationary model with a quasi-inflection point is plotted in the right panel of Fig. \ref{fig:Modes}. Importantly, the exponential amplification of the curvature fluctuations in the regime beyond SR with $3-\eps+\epss<0$ produces a peak in $\Xi_{\phi\phi}$ with respect to the usual SR result where $\Xi^{_\text{SR}}_{\phi\phi}=H^2/4\pi^2$.

One important remark about the evolution of the quantum fluctuations is that they depend on the dynamics of the coarse-grained fields through $\epsn$ that depend on $\pip$ and the potential $V$ that depends on $\bphi$. Therefore, the noise depends on the coarse-grained fields too. This implies that one would need to solve simultaneously the stochastic evolution dictated by the Langevin equations with the computation of the noise, which is a rather involved numerical problem. 
 One way to proceed would be through an iterative process in which the noise term generated by quantum fluctuations will back-react on the classical background trajectory. This new background will then be used to compute the stochastic-corrected quantum noise and thus iteratively generate the coarse-grained fields. We expect this iterative process to converge rapidly. As a first approximation, we computed the noise from the quantum fluctuations over the classical trajectory and then solved the stochastic equations. We have checked that the correction due to the first iteration is already subdominant, justifying this approximation. In the next section we will detail this computation. 

%-----
%SEC 4: N-POINT CORRELATION
%-----
\section{Correlation functions}
\label{sec:Correlation functions}

From the system of Langevin equations for the coarse-grained fields $\bphi$ and $\pip$ we can obtain the associated Fokker-Planck equation\footnote{One should remember that the connection between the Langevin and Fokker-Planck equations exists because the noise $\Xi_{AB}$ is white and Gaussian (see for instance a derivation in App. B of \cite{Tolley:2008na}).} for the probability density function (PDF) $P(N;\bPhi)$
\be
\label{eq:Fokker-Planck}
\frac{dP(N;\bPhi)}{dN}=-\frac{\partial}{\partial \Phi_A}\lp D_{A}\cdot P(N;\bPhi)-\frac{\Xi_{AB}}{2}\frac{\partial P(N;\bPhi)}{\partial\Phi_B}\rp\,,
\ee
where $\bPhi$ is the vector field of $\bphi$ and $\pip$, and the indices $A,\,B$ sum over $\bphi$ and $\pip$. The vector $\textbf{D}$ represent the drift with components
\begin{align}
D_\phi&=\pip\,, \\
D_\pi&=-(3-\epsilon)\lp\pip+\kappa^{-2}(\ln V)_{,\phi}\rp\,,
\end{align}
which follow from the Langevin equation. We will treat $\Xi_{AB}$ as a function of time only. 

We are interested in computing the effects of the noise on the classical trajectory. For that purpose, we can define the \emph{stochastic fluctuation} as the difference of the coarse-grained field with respect to the classical trajectory
\begin{align}
&\dphi=\bphi-\phi_c \\
&\dpi=\pip-\pi_c\,, 
\end{align}
and solve their evolution. Since the Fokker-Planck equation is hard to solve numerically, we can rewrite the problem in terms of the statistical moments of the PDF $\mpp{m}{n}$. A general correlation function of the stochastic fluctuations is defined as
\be
\mpp{n}{m}(N)=\int d\pip\int d\bphi\,(\bphi-\phi_c(N))^n(\pip-\pi_c(N))^m P(N;\bphi,\pip)\,.
\ee
Taking a time derivate of $\mpp{m}{n}$ and substituting the Fokker-Planck equation (\ref{eq:Fokker-Planck}), we obtain the general equation dictating its evolution
\begin{align}
\frac{d\mpp{n}{m}}{dN}&=n\lp\mppp{n-1}{m}{D_\phi}-\mpp{n-1}{m}D_\phi^c\rp+m\lp\mppp{n}{m-1}{D_\pi}-\mpp{n}{m-1}D_\pi^c\rp \nonumber \\
&+\frac{1}{2}n(n-1)\Xi_{\phi\phi}\mpp{n-2}{m}+\frac{1}{2}m(m-1)\Xi_{\pi\pi}\mpp{n}{m-2} \nonumber \\
&+\frac{1}{2}n\,m\lp\Xi_{\phi\pi}+\Xi_{\pi\phi}\rp\mpp{n-1}{m-1}\,, \label{eq:dmpp}
\end{align}
where $D_\phi^c$ and $D_\pi^c$ are the drift terms evaluated at the classical trajectory. For generic drift functions $D_\phi$ and $D_\pi$, we can obtain their statistical average expanding in powers of $\delta\phi$ and $\delta\pi$. In this sense, the terms $D_\phi^c$ and $D_\pi^c$ will be killing the first term of the expansion. Another interesting point is that only the symmetric part of the noise correlation $\Xi$ enters in the equations. In Appendix \ref{app:details} we provide the detailed form of this equation expanding the drift terms.

In order to obtain the PDF exactly, one would need to solve the infinite system of first-order differential equations described by (\ref{eq:dmpp}). In practice, one can truncate the series at a given order in which the higher moments are subdominant. The fact that we are solving first-order differential equations speeds the numerical analysis. Moreover, truncating at a given order $n$, we can always solve the evolution of the statistical moments of order $n$ which will only depend of the others of the same order. This serves to obtain the leading contribution to each order. We will discuss this in the next subsection. Later, we will comment on how to reconstruct the PDF from the $n$-point correlation functions.

%Power Spectrum
\subsection{Power Spectrum}

The basic observable from inflation is the power spectrum of scalar perturbations. In order to obtain it, we have to compute the 2-point correlation functions. At lowest oder, their evolution is described by the following system of equations
\be
\label{eq:system2}
\frac{d}{dN}\bpm \mphi{2} \\ \mphp \\ \mpi{2} \epm= \bpm 0 & -2 & 0 \\ g(\epsn) & -f(\epsn) & -1 \\ 0 & 2g(\epsn) & -2f(\epsn) \epm \bpm \mphi{2} \\ \mphp \\ \mpi{2} \epm +\bpm \Xi_{\phi\phi} \\ \Xi^s_{\phi\pi} \\ \Xi_{\pi\pi} \epm
\ee
where we have defined
\begin{align}
f(\epsn)&\equiv3(1-\eps)+(\ln V)_\phi\pi_c=3-\eps\lp1-\frac{\epss}{3-\eps}\rp\,, \\
g(\epsn)&\equiv\frac{(3-\eps)}{\kappa^2}(\ln V)_{\phi\phi}=-\frac{\epss}{2}\lp f(\epsn)+\frac{1}{2}\epss+\epsss\rp\,,
\end{align}
and $\Xi^s_{\phi\pi}$ denotes the symmetrization of the non-diagonal noise, i.e. $\Xi^s_{\phi\pi}=(\Xi_{\phi\pi}+\Xi_{\pi\phi})/2$. Once we have solved these equations, we can compute the power spectrum noting that the stochastic curvature perturbation $\zeta$, in our gauge choice\footnote{We remind the reader that this is because we have already incorporated the curvature fluctuations in the effective, time-dependent mass of $\delta\phi$ (see discussion in Sec. \ref{sec:Stochastic}).}, follows from  
\be
\langle \zeta^2\rangle=\frac{\kappa^2}{2}\frac{\mphi{2}}{\eps}\,.
\ee
Consequently, we can obtain the power spectrum from
\be
\label{eq:powespectrum}
\PS_\zeta=\frac{d\langle\zeta^2\rangle}{d\ln k}=\frac{1}{1-\eps}\frac{d\langle\zeta^2\rangle}{dN}=\frac{1}{1-\eps}\frac{\kappa^2}{2\eps}\lp\frac{d\langle\delta\phi^2\rangle}{dN}-\epss\langle\delta\phi^2\rangle\rp\,.
\ee
If we now introduce the equations for the 2-point function, we get
\be
\PS_\zeta=\frac{1}{1-\eps}\frac{\kappa^2}{2\eps}\lp\Xi_{\phi\phi}-2\mphp-\epss\mphi{2}\rp\,.
\ee
where
\be
\label{eq:phipi}
\mphp=\frac{f\,g}{f^2+g}\mphi{2}+\frac{1}{2(f^2+g)}\lp\frac{d\mpi{2}}{dN}-\Xi_{\pi\pi}-2f\lp\frac{d\mphp}{dN}-\Xi_{\phi\pi}\rp\rp\,.
\ee
Noticeably, the power spectrum does not only depend on $\mphi{2}$ and $\Xi_{\phi\phi}$. Whenever we are in the SR regime, the derivatives of $\mphp$ and $\mpi{2}$ can be neglected as well as their associated noise terms. At leading order in the SR parameters, we have that $f\approx3$ and $g\approx-3\epss/2$. This means that $\mphp\approx-\epss/2$ and thus we recover the usual expression for the power spectrum
\be
\PS_\zeta\simeq\frac{\kappa^2}{2\eps}\Xi_{\phi\phi}\,.
\ee
Note that if the noise is properly computed accounting for the growth after horizon crossing as previously discussed, this power spectrum is equivalent to the standard Mukhanov-Sasaki formalism. This result confirms that, at leading order, SR stochastic inflation gives the same results as the curved space, quantum field theory techniques \cite{Vennin:2015hra}. For completeness, we derive this result also directly from SR stochastic inflation in Appendix \ref{app:StochasticSR}.

However, whenever we are beyond SR, there could be additional contributions to the power spectrum. This is similar to the case of hybrid inflation, were the diffusion due to the second field can significantly amplify the power spectrum \cite{Clesse:2015wea}. For the case of single field inflation with a quasi-inflection point, any deviation would be around this critical point. Since (\ref{eq:system2}) is a system of first order differential equations with non-constant coefficients there is no generic analytical solution. We will thus solve the system numerically. 

One should note that whenever the quantum fluctuations are small, the combination of $-2\mphp-\epss\mphi{2}$ is subdominant with respect to $\Xi_{\phi\phi}$ and therefore, the usual expression for $\PS_\zeta$ in SR is recovered, cf. (\ref{eq:powespectrum}). However, since in our case short wavelength fluctuations become large, introducing a peak in the noise, they also produce a peak in $\mphi{2}$ and $\mphp$ and their contribution is no longer negligible. As a consequence, there will be a peak in the power spectrum produced by the diffusion. This peak will coincide with the maximum of the function multiplying $\mphi{2}$, i.e. $\frac{-2f\,g}{f^2+g}-\epss$. In fact, around this peak the power spectrum can be approximated by
\be
\PS_\zeta\approx\frac{\kappa^2}{2\eps}\lp\Xi_{\phi\phi}-c\,\epss\rp\,,
\ee
where $c$ is a constant. This is because $\frac{-2f\,g}{f^2+g}-\epss\approx-\epss$, $\mphi{2}\approx c$ and the contribution of the other terms in (\ref{eq:phipi}) is subdominant. Then, the location of the diffusion peak $N_{\mathrm{max}}$ is given by the maximum of $\epss/\eps$, which can be solved from
\be
\epsss(N_{\mathrm{max}})=\epss(N_{\mathrm{max}})\,.
\ee 
In addition, we find that the relative height of the diffusion peak compared to the standard peak in $\PS_\zeta$ scales with the size of the peak of the noise $\Xi_{\phi\phi}^\text{max}$. This is sensitive to the transition between the first and second plateau of the potential or, in other words, to the ratio between the maximum and the minimum of $\eps$. This tells us that the more the inflaton slows down approaching the quasi-inflection point, the higher the diffusion peak will be. Importantly, this enhancement of the power spectrum by quantum diffusion emerges on top of the peak due to the classical dynamics. This implies that one obtains much higher amplitudes in the power spectrum than expected with the standard techniques.

\begin{figure}[!t]
\centering
\includegraphics[width = 0.8\textwidth]{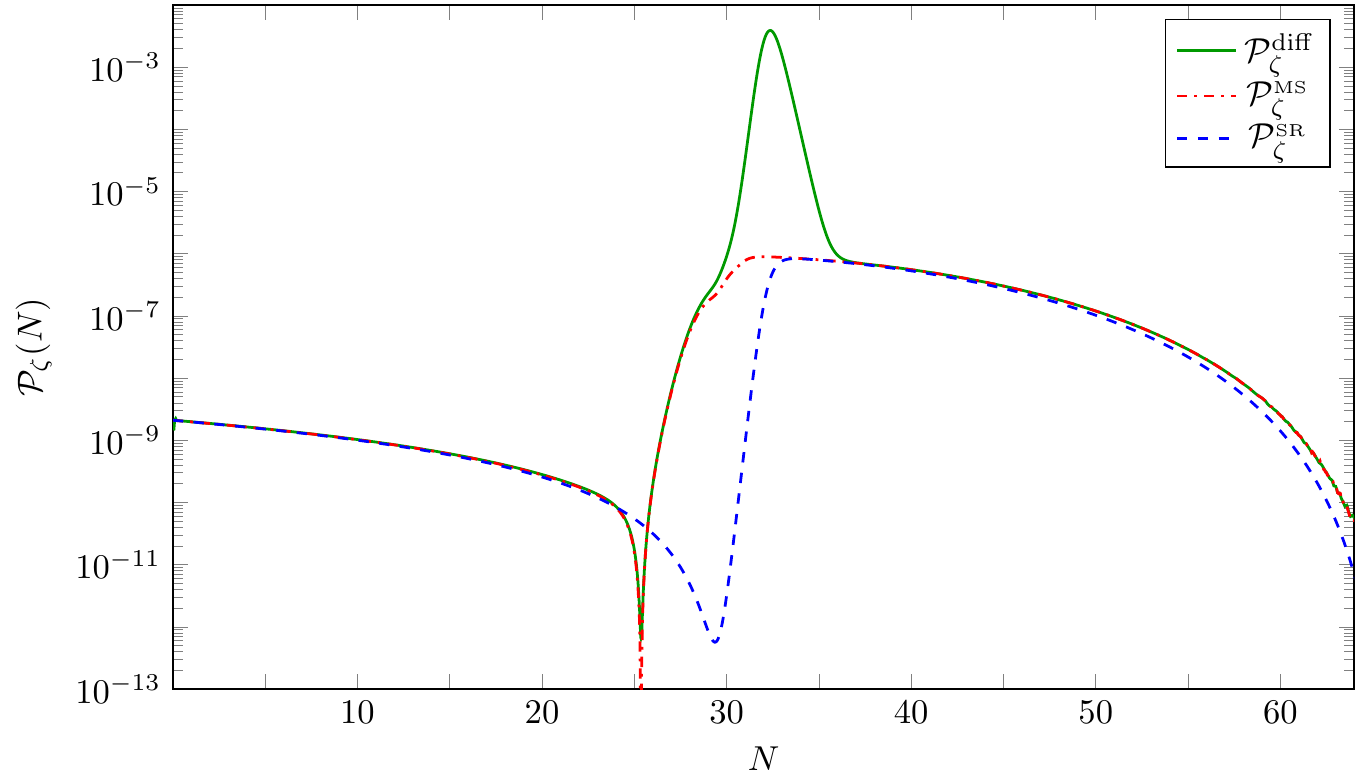}
\vspace{-5pt}
\caption{Power spectrum for the curvature perturbations $\PS_\zeta$ computed from quantum diffusion (solid green), Eq. (\ref{eq:powespectrum}), solving the Mukhanov-Sasaki equation (red dash-dotted), Eq. (\ref{eq:PMS}), and using the slow-roll formula (blue dashed) in Eq. (\ref{eq:PSR}).}
\label{fig:PowerSpectrum}
\end{figure}

In Fig. \ref{fig:PowerSpectrum}, we present power spectrum for the curvature fluctuations for our inflationary model with a quasi-inflection point resulting from stochastic inflation beyond SR, Eq. (\ref{eq:powespectrum}), the Mukhanov-Sasaki equation, Eq. (\ref{eq:PMS}), and the SR approximation, Eq. (\ref{eq:PSR}). There are different comments to place in order. Firstly, we observe that there are differences between the numerical solution of $\zeta_k$ and the SR approximation at large scales. The difference around $N\sim30$ are produced for the already discussed growth of the perturbations after horizon crossing. Interestingly, this shifts the raising of $P_\zeta$ to higher scales. Depending on the size of the peak, this could imply approaching to the detectability region of spectral distortions. Secondly, there are also differences with the SR approximation at the end of inflation. This is because $\epss$ is of order 1 during this period and the scalar, metric fluctuations $\Phi$ are excited. For certain shapes of the potential at the end of inflation, this difference could be significantly larger. Finally, we observe the peak resulting from quantum diffusion beyond SR discussed before. This large enhancement of the fluctuations will certainly have relevant implications for the production of PBHs.

% FIG. 3
%

%--------
%NON-GAUSSIANITIES
%--------
\subsection{Non-Gaussianities}

One of the nice things of stochastic inflation is that it allows to compute higher order correlators of the coarse-grained field. The fact that there is a peak in the noise will also affect the non-Gaussian contributions to the PDF. We expect that any significant non-Gaussianity is localized around this region of large stochastic noise. We can solve any $n$-point function using the system of equations (\ref{eq:dmpp}). In analogy to the second moments, the leading contribution to the 3-point functions is
\be
\label{eq:system3}
\frac{d}{dN}\bpm \mphi{3} \\ \mphinpi{2} \\ \mphipin{2} \\ \mpi{3} \epm= \bpm 0 & -3 & 0 & 0 \\ g(\epsn) & -f(\epsn) & -2 & 0 \\ 0 & 2g(\epsn) & -2f(\epsn) & -1 \\ 0 & 0 & 3g(\epsn) & -3f(\epsn) \epm \bpm \mphi{3} \\ \mphinpi{2} \\ \mphipin{2} \\ \mpi{3} \epm +\bpm 3\Xi_{\phi\phi}\mph \\ \Xi_{\phi\phi}\mpii+\Xi^s_{\phi\pi}\mph \\ \Xi_{\pi\pi}\mph+\Xi^s_{\pi\phi}\mpii \\ 3\Xi_{\pi\pi}\mpii \epm\,.
\ee
A similar expression can be obtained for the 4-point function.

For the purpose of our analysis, we are interested in the Fourier transform of the $n$-point functions. In particular, we are interested in the 3 and 4-point functions
\begin{align}
&\langle\zeta^3_k\rangle=\frac{\kappa^{3}}{2^{3/2}}\frac{d^2}{d\ln k^2}\lp\frac{\mphi{3}}{\eps^{3/2}}\rp & &\text{and} & &\langle\zeta^4_k\rangle=\frac{\kappa^4}{4}\frac{d^3}{d\ln k^3}\lp\frac{\mphi{4}}{\eps^2}\rp\,\,
\end{align}
which are associated respectively to the bispectrum and the trispectrum. Within this notation, $\langle\zeta^2_k\rangle=\PS_\zeta$. To measure the relative strength of these contributions one can normalize them with respect to the power spectrum, i.e. $\langle\bar{\zeta^3_k}\rangle=\langle\zeta^3_k\rangle/\langle\zeta^2_k\rangle^{3/2}$ and $\langle\bar{\zeta^4_k}\rangle=\langle\zeta^4_k\rangle/\langle\zeta^2_k\rangle^{2}$. In fact, these normalized moments will be the relevant quantities when computing the probability to form a PBH.

%--------
%CHARACTERISTIC FUNCTION
%--------
\subsection{Characteristic function}
\label{sec:CharacteristicFunction}

In order to connect the $n$-point correlators with the PDF we can introduce the characteristic function $\chi(u_\phi,u_\pi)$, defined as the Fourier transform of the PDF
\be
\chi(u_\phi,u_\pi)=\int_{-\infty}^\infty d\dphi\int_{-\infty}^\infty d\dpi e^{i(u_\phi\dphi+u_\pi\dpi)}P(N;\dphi,\dpi)\,,
\ee
where we are introducing now the PDF of the stochastic fluctuations $\dphi$ and $\dpi$. By denoting $\bu=(u_\phi,u_\pi)$ and $\bx=(\dphi,\dpi)$ we recover the generic multivariate definition of $\chi(\bu)=\langle e^{i\bu\cdot\bx}\rangle$. The characteristic function is generated through the cumulant tensors $\kappa_{i_1\cdots i_n}$ defined by
\be
\kappa_{i_1\cdots i_n}=(-i)^n\left.\frac{\partial}{\partial u_1}\cdots \frac{\partial}{\partial u_n}\log\langle e^{i\bu\cdot\bx}\rangle\right\vert_{\bu=0}\,.
\ee
In our case, the indices $i_n$ could take only two values, either $\phi$ or $\pi$. 

\begin{figure}[!t]
\centering
\includegraphics[width = 0.478\textwidth]{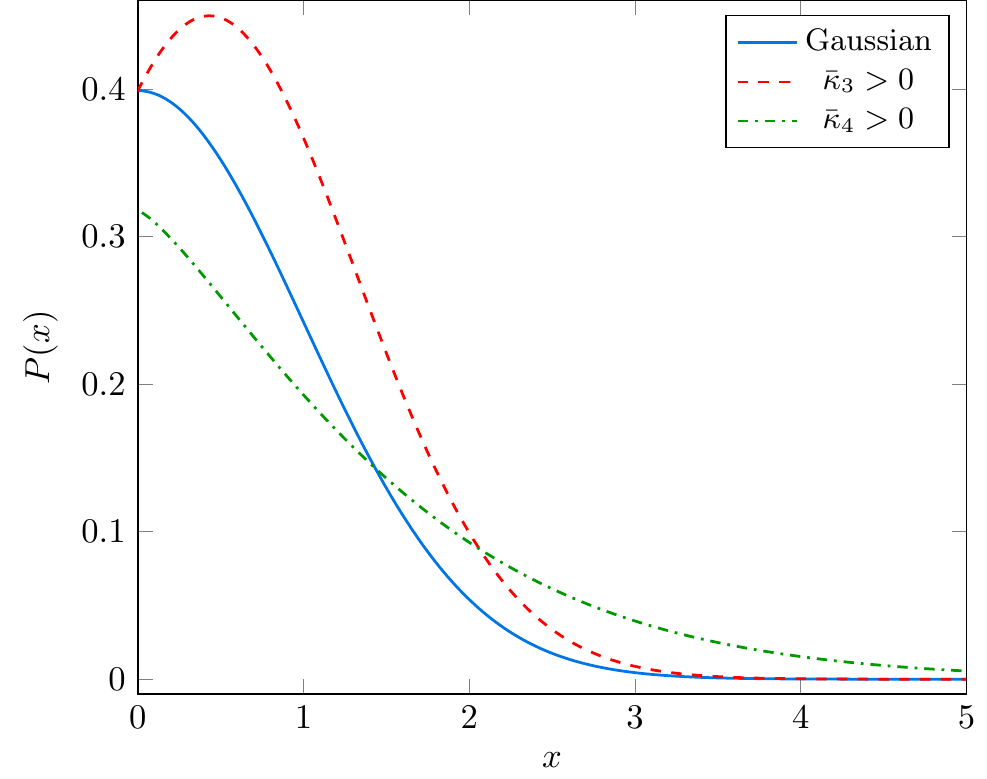}
\includegraphics[width = 0.49\textwidth]{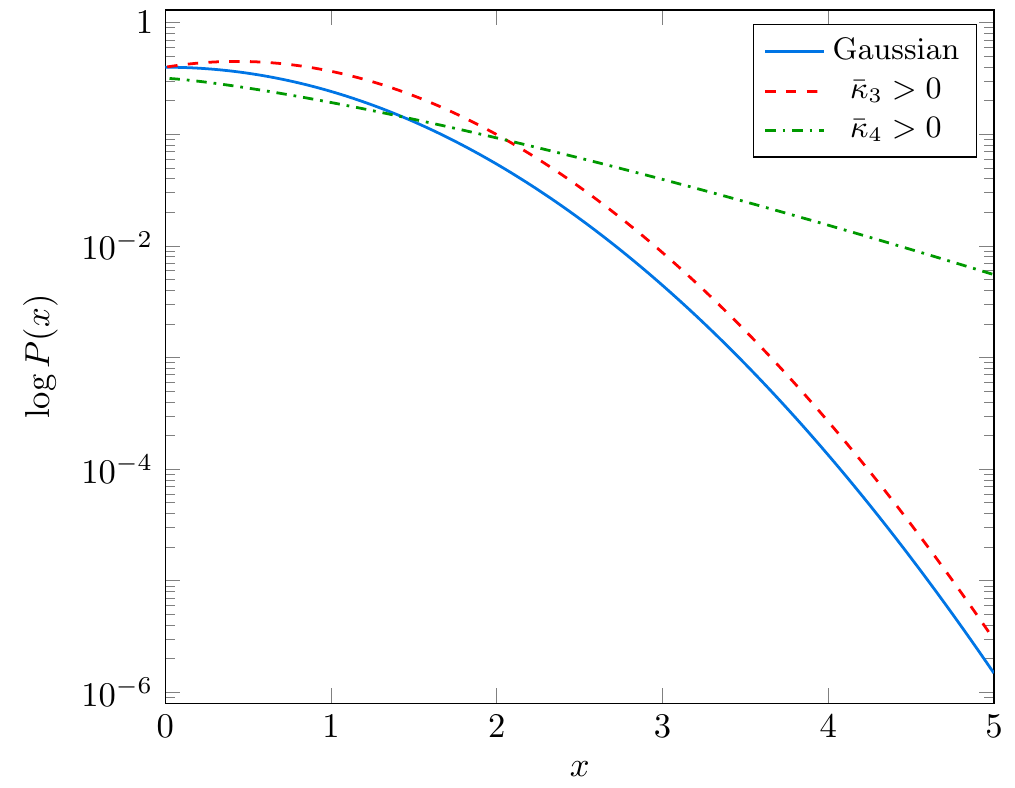}
\vspace{-5pt}
\caption{Illustration of the effect of a positive skewness, $\bar{\kappa}_3$, and a positive (excess) kurtosis, $\bar{\kappa}_4$, for the probability density function $P(x)$ (left) and its logarithm (right) in comparison with a Gaussian distribution ($\kappa_{n>2}=0$). The dominant effect in the tail is given by a positive kurtosis.}
\label{fig:PDF}
\end{figure}

The cumulant tensors can be generically related to the $n$-point correlation functions. However, as it will be clarified in the next section, we will only be interested in the case in which we marginalize over the momentum. For the generic $n$-variate case, this is equivalent to marginalize over $n-1$ variables, which effectively leaves a univariate distribution with characteristic function $\chi(u_1)=\left.\chi(\bu)\right\vert_{u_{i>1}=0}$. For our problem, we will have 
\be
\label{eq:SingleCharacteristicFunction}
\chi(u_\phi)=\langle e^{iu_\phi\dphi}\rangle=\exp\lb{\sum_{p=1}^{\infty}\frac{i^p}{p!}}\kappa_{p}u_\phi^{p}\rb\,,
\ee
where the first cumulants are given by
\be
\begin{split}
\kappa_1&=\mph \\
\kappa_2&=\mphi{2}-\mph^{2} \\
\kappa_3&=\mphi{3}-3\mph\mphi{2}+2\mph^{3} \\
\kappa_4&=\mphi{4}-4\mph\mphi{3}+12\mph^2\mphi{2}-3\mphi{2}^2-6\mph^4\,.
\end{split}
\ee
With all these statistical machinery, we are now ready to compute the probability to form a PBH. Before doing so, let us briefly review the effect of the third and fourth moments in the PDF. For illustrative purposes, we have compared a Gaussian distribution ($\kappa_{n>2}=0$) to a PDF with $\kappa_3,\kappa_4\neq0$ in the left panel of Fig. \ref{fig:PDF} (see specific formula we are plotting in (\ref{eq:generalPDF})). On the one hand, the third moment can induce an asymmetry of the PDF with respect to its mean value. This is commonly characterized by the skewness, which in terms of the cumulants reads $\bar{\kappa}_3=\kappa_3/\kappa_2^{3/2}$. On the other hand, the fourth moment tell us how relevant are the tails with respect to the peak and it is measured by the (excess) kurtosis $\bar{\kappa}_4=\kappa_4/\kappa_2^2$. For us, since we are interested in the probability above a certain threshold, the important part will be the tails of the PDF that can be better seen plotting the logarithm of the PDF as in the right panel of Fig. \ref{fig:PDF}. There, one can clearly see that both a positive skewness and kurtosis induce higher tails compared to the Gaussian. For our toy model, we find that the tails of the PDF are dominated by a positive kurtosis. This will enlarge the probability to form a PBH. However, one should remember that non-Gaussianities could also reduce the fraction of PBH if the distribution is characterized by a negative skewness or kurtosis.

% FIG. 4
%

%--------------
%SECTION 5: IMPLICATIONS FOR PBH
%--------------
\section{Implications for PBH production}
\label{sec:PBH}

PBH are formed when large fluctuations produced during inflation re-enter the horizon and collapse all the content within a causal horizon. The fraction of PBH formed thus depends on the probability to have fluctuations larger than a certain threshold at the scale of the horizon. The probability distribution in single-field inflation is a bivariate distribution that depends on the fluctuations and their velocities. Therefore, the fraction of PBH formed is computed from
\be
\beta_f(M)=P(\zeta>\zeta_c)=\int_{\zeta_c}^{\infty}\int_{-\infty}^{\infty}P(M;\zeta,\dpi)d\delta\pi d\zeta\,,
\ee
where we have integrated over the momentum fluctuations.\footnote{One may wonder if the gravitational collapse could depend on the canonical momentum fluctuations $\delta\pi$, which could happen if entropy as well as adiabatic density fluctuations existed at horizon re-entry.
However, although entropy fluctuations can be excited during the non-slow-roll phase, only adiabatic perturbations are present at the end of inflation. This means that only curvature fluctuations will re-enter the horizon and, as a consequence, whether PBH are formed or not depends only on the size of curvature gradients. That is the reason why we must marginalize over the canonical momentum fluctuations.
Another issue might be the effect of non-Gaussianities in the formation of PBH, which has not been studied yet in depth in the literature \cite{Yoo:2018kvb}, and would require a dedicated study.} 
Here, we have expressed the PDF in terms of the mass $M$ of the PBHs formed instead of the number of $e$-folds. This is because the mass of the PBHs is related to the size of the causal horizon collapsing that, at the same time, depends on the time of formation (see e.g. \cite{GarciaBellido:1996qt})
\be
M_{_\text{PBH}}\simeq 4\pi\gamma \frac{M_{\text{pl}}^2}{H_{\text{inf}}}e^{2N}\,,
\ee
where $M_{\text{pl}}$ is the reduced Planck mass, $H_{\text{inf}}$ the energy scale of inflation and $\gamma$ efficiency parameter encapsulating the details of the gravitational collapse and the efficiency of reheating, that we fix to be $\gamma\sim0.2$. Finally, let us note that we can directly translate the PDF of the inflaton fluctuations $\dphi$ to the PDF of curvature perturbations $\zeta$ since both perturbations are related by a classical field $z$.

With the fraction of PBHs forming $\beta_f(M)$ one can compute the contribution of PBHs to the energy density of the universe and determine the fraction of the DM that they represent. In this sense, it is convenient to compute this at the time of matter-radiation equation
\be
\Omega_{_\text{PBH}}^\text{eq}=\int_{M_\text{ev}}^{M_\text{eq}}\beta_\text{eq}(M)d\ln M
\ee
where $M_\text{ev}$ is the mass threshold for Hawking evaporation and we have assumed that the fraction of PBH at the time of equality has grown due to cosmic expansion from the time of formation $\beta_\text{eq}(M)=e^{(N_\text{eq}-N_f)}\beta_f(M)$. 

Altogether, the abundance of PBHs will be sensitive to the PDF of the stochastic fluctuations. We will show next how this abundance can be connected with the correlation functions of these fluctuations. For the case of Gaussian fluctuations then all the information is contained in the 2-point functions that can be related to the power spectrum $P_\zeta$. A peak in $P_\zeta$ thus produce a higher probability to form PBHs. This is the whole purpose of considering inflation with a quasi-inflection point as a source of PBHs. However, if the fluctuations are non-Gaussian (as it will be in our case), these additional contributions from higher $n$-point functions will be relevant. This is because we are interested in the probability of being above a certain threshold and, as a consequence, we are very sensitive to the tails of the distribution. The fact that non-Gaussianities can be relevant in the production of PBHs has already been considered in \cite{Saito:2008em,Byrnes:2012yx,Young:2013oia,Young:2015cyn,Tada:2015noa,Young:2015kda,Franciolini:2018vbk}. When considering quantum diffusion in SR inflation, non-Gaussianities played a central role \cite{Bullock:1996at,Ivanov:1997ia,Yokoyama:1998pt,Pattison:2017mbe}. Here, for quantum diffusion beyond SR, non-Gaussianities will as well be very important.

%ABUNDANCE
\subsection{Abundance of PBH}

The abundance of PBH is then determined by $P(\zeta>\zeta_c)=\langle\Theta(\zeta-\zeta_c)\rangle_{_P}$. We can express this probability in terms of the characteristic function by
\be
\label{eq:ProbabilityAbove}
P(\zeta>\zeta_c)=\int_{-\infty}^\infty d\zeta\int_{\zeta_c}^{\infty}da\int_{-\infty}^\infty\frac{du}{2\pi}e^{iu(\zeta-a)}P(\zeta)=\int_{\zeta_c}^{\infty}da\int_{-\infty}^\infty\frac{du}{2\pi}e^{-iua}\chi(u)\,,
\ee
where in the first equality we have introduce the integral definition of the step function\footnote{The step function can be written as $\Theta(x-x_c)=\int_{x_c}^{\infty}da\int_{-\infty}^\infty\frac{du}{2\pi}e^{iu(x-a)}$.}. Therefore, as anticipated in Sec. \ref{sec:CharacteristicFunction}, this probability is only sensitive to the marginalized characteristic function $\chi(u)$. Such function can be constructed with the cumulants of the PDF, cf. (\ref{eq:SingleCharacteristicFunction}), which can be obtained solving the evolution of the correlations functions (\ref{eq:dmpp}) as explained before.

Conveniently, the integral of $P(\zeta>\zeta_c)$ can be solved exactly. The resulting formula is
\be
\label{eq:ProbabilityPBH}
P(\zeta>\zeta_c)=\frac{1}{2}\text{Erfc}\lb\frac{z_c}{\sqrt{2}}\rb+\frac{e^{-z_c^2/2}}{\sqrt{\pi}}\sum_{n=3}^{\infty}\frac{\bar{K}_n}{2^{n/2}n!}H_{n-1}\lb\frac{z_c}{\sqrt{2}}\rb\,,
\ee
where $\text{Erfc}$ is the complementary error function and $H_n$ are the Hermite polynomial. The normalized threshold is given by $z_c=(\zeta_c-\kappa_1)/\kappa_2^{1/2}$, where we recall that $\kappa_n$ are the cumulants of the PDF. The functions $\bar{K}_n$ can be related to the normalized cumulants $\bar{\kappa}_n$ and only differ from them for $n\geq6$.
This result was originally obtained in Ref. \cite{Matarrese:1986et} using the path integral formalism and recently revisited in \cite{Franciolini:2018vbk}. Since it is an important formula, we summarize for completeness the derivation of this result and the details in the definitions in App. \ref{app:statistics} using the language of the characteristic function and cumulants that we are using here.

\begin{figure}[!t]
\centering
\includegraphics[width = 0.47\textwidth]{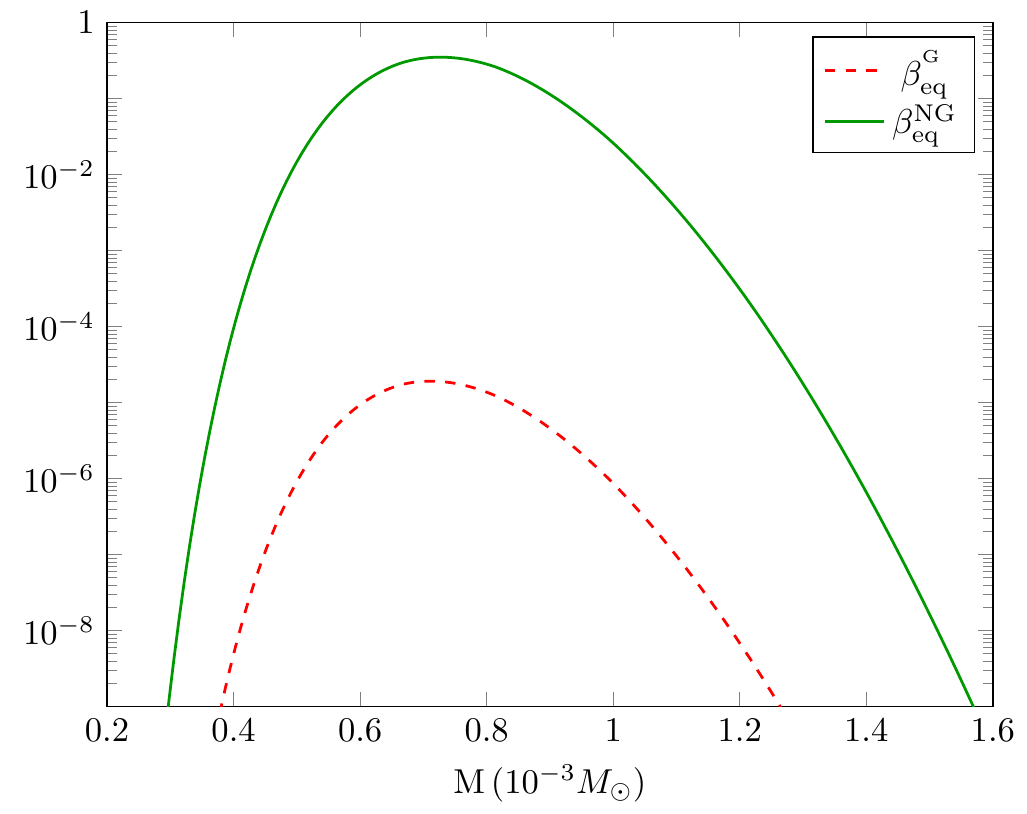}
\vspace{-5pt}
\caption{Fraction of PBH at the time of equality $\beta_\text{eq}$ as a function their mass computed using quantum diffusion and fixing $\zeta_c=0.5$. The abundance given by the Gaussian approximation is compared with the non-Gaussian, which is orders of magnitude larger.}
\label{fig:AbundancePBH}
\end{figure}

Regarding the effects of quantum diffusion on the abundance of PBH, there will be two effects. First, the fact that the slow-roll violation enhances quantum fluctuations that back-react producing an additional peak (on top of the classical one) in the power spectrum will significantly increase the number of PBH that are formed. This can be very easily understood looking at the first term of (\ref{eq:ProbabilityPBH}), which accounts for the Gaussian contribution. The threshold $z^\text{diff}_c$ for quantum diffusion will be lowered with respect to the usual one $z^{_\text{MS}}_c$ by
\be
z^\text{diff}_c\simeq\lp\frac{\PS_\zeta^{_\text{MS}}}{\PS_\zeta^{_\text{diff}}}\rp^{1/2}z^{_\text{MS}}_c\,.
\ee
Depending on the size of the diffusion peak, the difference around this point could be of several orders of magnitude. In Fig. \ref{fig:AbundancePBH}, we present the fraction of PBH at equality $\beta^{_\text{G}}_\text{eq}$ for a Gaussian distribution with variance given by the power spectrum computed using stochastic inflation $(\PS_\zeta^\text{diff})^{1/2}$ (cf. Fig. \ref{fig:PowerSpectrum}) and $\zeta_c=0.5$. The corresponding fraction of PBH without considering the diffusion, computed from $(\PS_\zeta^{_\text{MS}})^{1/2}$, is very suppressed, due to several orders of magnitude difference in the peak, and does not appear in the plot.

% FIG. 5

Second, non-Gaussiniaties will change the above prediction by either augmenting or decreasing the number of PBH depending on the sign of each contribution. For our case of study, we solve the stochastic evolution up to fourth order.\footnote{We have checked that the fifth order contribution is negligible. However, we realize that in order to derive the sixth order reliably we require higher time resolution than the one used in this analysis.} We find that the fourth moment is the dominant piece. It contributes with a positive large kurtosis. This significantly amplifies the production of PBHs as it can be seen from the fraction of PBH at equality $\beta^{_\text{NG}}_\text{eq}$ plotted in Fig. \ref{fig:AbundancePBH}. In this example, we have chosen the parameters so that the energy density of PBH at equality $\Omega_{_\text{PBH}}^\text{eq}$, when considering the NG contributions, is the dominant fraction of the DM. Considering only the Gaussian contribution would lead to PBHs being a very subdominant fraction of DM. This shows that indeed NG are very relevant and they could soften the necessity to have very large peaks in the power spectrum. In the following, we discuss the implications of these results for inflationary model building and how they generalize.

%MODEL BUILDING
\subsection{Model building}

In the previous section, we have found that in models of inflation with a quasi-inflection point quantum diffusion causes two main effects: $i)$ there is an enhancement of the power spectrum and $ii)$ non-Gaussian corrections become relevant, modifying the prediction for the abundance of PBHs. These results have profound implications when constructing models to produce PBHs. On the one side, there will be regions of the parameter space of previously considered models in the literature now excluded because they would overproduce PBHs. On the other side, there will be other regions, possibly with less tuning of the parameters, that now would copiously produce PBHs.

For the purpose of this work, we have not performed a detailed analysis of the viable parameter space for inflationary models with a near-inflection point accounting for quantum diffusion. Such re-analysis should be performed elsewhere. Our expectation is that any quantum diffusion effect would be closely related to the type of deviation from slow-roll. For example, we find that the height of the diffusion peak is linked to how much the inflaton slows-down, i.e. how small $\eps$ becomes, while its width depends on how fast the transition from slow-roll to ultra slow-roll happens, i.e. for how long $\epss<0$. In any case, we observe that the violation of SR should be such that during an interval $3-\eps+\epss<0$ so that there is an amplification of the stochastic noise. Moreover, we encounter that the enhancement of the power spectrum and the relevance of non-Gaussian corrections are highly correlated. This implies that the larger the diffusion peak is, the more the Gaussian prediction for the abundance of PBHs will be modified. For the toy model potential that we have considered, we find that a positive kurtosis dominates over the skewness to enlarge the production of PBHs. From these considerations, we expect that quantum diffusion would have an important role in models such a critical Higgs inflation \cite{Ezquiaga:2017fvi}, single-field double inflation \cite{Kannike:2017bxn}, radiative plateau inflation \cite{Ballesteros:2017fsr} or some string inspired realizations \cite{Cicoli:2018asa,Ozsoy:2018flq,Dalianis:2018frf}.

%--------------
%SEC 6: CONCLUSIONS
%--------------
\section{Conclusions}
\label{sec:Conclusions}

It is interesting to think about the possibility that there is a large fraction of primordial black-hole in the universe since they could help in solving some of the problems of modern cosmology such as the DM \cite{Garcia-Bellido:2017fdg} or the origin of supermassive BHs \cite{Clesse:2015wea,Carr:2018rid}. Here, we have considered how PBHs are produced in single field models of inflation with a quasi-inflection point. We have taken into account the effect that quantum fluctuations have on the inflationary dynamics. For that, we have used the tools of stochastic inflation. Since the production of PBHs is related to a deviation of SR, we have considered the system of Langevin equations associated to both the coarse-grained field $\bphi$ and momentum $\pip$ in the Hamilton-Jacobi formalism. In order to find the multi-variate probability density function associated to this stochastic problem, we have solved the system of first-order differential equations for the statistical moments $\mpp{m}{n}$. With these moments one can construct the characteristic function $\chi(u_\phi,u_\pi)$ that can be used to compute the PDF. The methodology that we have developed can be applied to study quantum diffusion effect beyond slow-roll in general. We have particularized its use to find the probability that a given fluctuation is above a certain threshold $P(\zeta>\zeta_c)$, which is the relevant question to know if PBHs are formed.

We have analyzed the implications of quantum diffusion beyond SR for the production of PBHs. Inflationary models with a quasi-inflection point induce an exponential growth of the curvature modes whenever $3-\eps+\epss<0$. We find that this enhancement of the quantum fluctuations also enlarges the stochastic noise. In turn, this produces a peak in the power spectrum on top of the spectrum of $\zeta_k$ without diffusion. Such a diffusion peak has a clear effect amplifying the PBH production. Moreover, we observe that this large stochastic noise also induce important non-Gaussian contributions. These NG contributions could either enlarge or decrease the PBH production. For our toy model potential, we realize that the dominant contribution is given by the fourth moment. It introduces a positive contribution that augments the weight of the tails of the PDF, leading to a significantly higher probability to form PBHs. Adding both contributions, we find that for our case of study quantum diffusion substantially boosts the formation of PBHs.

Our results show that quantum diffusion beyond SR can have an important role in determining the abundance of PBHs for single-field models with a quasi-inflection point. These effects can be parametrized with the dynamics of the Hubble-flow parameters $\epsn(N)$. We expect that our conclusions could be extended to different potentials studied in the literature that share a similar behavior of $\epsn$. We believe that the predictions for the spectrum of PBHs in this class of models should be revisited in the future accounting for quantum diffusion.

%--------
%ACKNOWLEDGMENTS
%--------
\acknowledgments

We are very grateful to Vincent Vennin for useful conversations and comments on the manuscript. The authors acknowledge support from the Research Project FPA2015-68048-03-3P (MINECO-FEDER) and the Centro de Excelencia Severo Ochoa Program SEV-2012-0249. JME is supported by the Spanish FPU Grant No. FPU14/01618. He thanks CERN Theory Division for hospitality during his stay there and the FPU program for financial support. JGB acknowledges support from the Salvador de Madarriaga program Ref. PRX17/00056.

%--------
%APPENDIX
%--------
\appendix
%-------------
%APP. 1: Computing the noise
%-------------
\section{Computing the noise}
\label{app:noise}
In this appendix we complement the content of Sec. \ref{sec:Stochastic} by detailing the computation of the noise. For this calculation, we follow the work of Ref. \cite{Grain:2017dqa}. The starting point is to apply the coarse-graining to the fields $\phi=\phi_s+\bphi$ and $\pi_\phi=(\pi_\phi)_s+\pip$ through
\begin{align}
\phi_s&=-\int\frac{d^3k}{(2\pi)^{3/2}} W\lp\frac{k}{k_\sigma}\rp\lb a_{\kv} \phi_{\kv} \, e^{-i\kv\xv}+a^{\dagger}_{\kv} \phi^{\star}_{\kv} \, e^{i\kv\xv}\rb\,,\\ 
 (\pi_\phi)_s&=-\int\frac{d^3k}{(2\pi)^{3/2}}W\lp\frac{k}{k_\sigma}\rp\lb a_{\kv} \pi_{\kv} \, e^{-i\kv\xv}+a^{\dagger}_{\kv} \pi^{\star}_{\kv} \, e^{i\kv\xv}\rb\,, 
\end{align}
where $W\lp k/k_\sigma\rp$ is a window function selecting the modes with $k\gg k_\sigma=\sigma aH$ for $\sigma\ll1$.
Then, the evolution equations lead to
\begin{align}
\frac{d\bphi}{dN}&=\pip+(\pi_\phi)_s-\frac{d\phi_s}{dN}\,, \\
\frac{d\pip}{dN}&=-(3-\epsilon)\pip-\frac{V_{,\phi}}{H^2}-\frac{d(\pi_\phi)_s}{dN}-(3-\epsilon)(\pi_\phi)_s-\lp\lp\frac{k}{aH}\rp^2+\frac{m_\text{eff}^2}{H^2}\rp\phi_s\,,
\end{align}
where we have linearized in the short wavelengths modes and $m_\text{eff}$ is the time dependent effective mass that can be related to $V_{,\phi\phi}$ and $\epsn$, cf. (\ref{eq:modepi}). Subsequently, using the mode equations for $\phi_k$ and $\pi_k$ (\ref{eq:modephi}-\ref{eq:modepi}), one can realize that when taking the time derivatives of $\phi_s$ and $(\pi_\phi)_s$ all the terms associated with them cancel in the equations of motion but the one associated with the time derivatives of the window function. These remaining terms are thus the ones defining the noise
\begin{align}
\xi_\phi(N)&=-\int\frac{d^3k}{(2\pi)^{3/2}}\frac{d}{dN}\lb W\lp\frac{k}{k_\sigma}\rp\rb\lb a_{\kv} \phi_{\kv} \, e^{-i\kv\xv}+a^{\dagger}_{\kv} \phi^{\star}_{\kv} \, e^{i\kv\xv}\rb\,,\\ 
\xi_\pi(N)&=-\int\frac{d^3k}{(2\pi)^{3/2}}\frac{d}{dN}\lb W\lp\frac{k}{k_\sigma}\rp\rb\lb a_{\kv} \pi_{\kv} \, e^{-i\kv\xv}+a^{\dagger}_{\kv} \pi^{\star}_{\kv} \, e^{i\kv\xv}\rb\,, 
\end{align}
For a Heaviside window function, $W(k/k_\sigma)=\Theta(k/k_\sigma-1)$, its derivative is a delta function $\frac{d}{dN}\lb W\lp\frac{k}{k_\sigma}\rp\rb=\delta(k-k_\sigma)$. This implies that the integrals can be solved. The noise is characterized by the two-point auto-correlations
\be
\Xi_{AB}(N,N')\equiv\langle0\vert\xi_A(N)\xi_B(N')\vert0\rangle=\frac{k_\sigma^3(N)}{2\pi^2}\frac{d\ln k_\sigma}{dN}A_{k_\sigma}B^\star_{k_\sigma}\cdot\delta(N-N')\,,
\ee
where $A,\,B$ could be either $\phi$ or $\pi$. The fact that the noise is uncorrelated at different times (white/Markovian noise) is related to the choice of window function. Since the derivative of the Heaviside window function is a Dirac distribution in momentum space, the product of the two noises evaluated at different times will lead to $\delta(k-k_\sigma(N))\delta(k-k_\sigma(N'))=\delta(N-N')$. Having a Gaussian, white noise allows to connect the Langevin equations with the Fokker-Planck as discussed in the main text.

%-------------
%APP. 2: FOKKER-PLANCK IN DETAIL
%-------------
\section{Correlation functions in detail}
\label{app:details}

In Sec. \ref{sec:Correlation functions}, we have explained how to rewrite the problem of solving the Langevin equations in terms of a set of first order differential equations for the statistical moments $\mpp{m}{n}$. In this appendix we provide details of some of the formulae we have used. Since we want to compute the $n$-point functions of the stochastic fluctuations, $\dphi=\bphi-\phi_c$ and $\dpi=\pip-\pi_c$, we have to rewrite the drift vector as
\begin{align}
D_\phi&=-(\dpi+\pi_c)\,, \\
D_\pi&=-\lp3-\frac{\kappa^2}{2}(\dpi+\pi_c)^2\rp(\dpi+\pi_c)+\lp3-\frac{\kappa^2}{2}(\dpi+\pi_c)^2\rp\frac{\left.(\ln V)_{,\phi}\right\vert_{\dphi+\phi_c}}{\kappa^2}\,.
\end{align}
If we substitute these expression into the evolution equations (\ref{eq:dmpp}), we obtain
\be
\begin{split}
\frac{d\mpp{n}{m}}{dN}&=-n\mpp{n-1}{m+1}-3m\mpp{n}{m} \\
&+m\frac{\kappa^2}{2}\lp\mpp{n}{m+2}+3\mpp{n}{m+1}\pi_c+3\mpp{n}{m}\pi_c^2\rp \\
&+m\frac{(3-\eps)}{\kappa^2}\sum_{i=1}\frac{(\ln V)^{(i+1)}}{i!}\mpp{n+i}{m-1} \\
&-\frac{m}{2}\lp\mppp{n}{m+1}{(\ln V)_{,\phi}}+2\mppp{n}{m}{(\ln V)_{,\phi}}\pi_c\rp \\
&+\frac{1}{2}n(n-1)\Xi_{\phi\phi}\mpp{n-2}{m}+\frac{1}{2}m(m-1)\Xi_{\pi\pi}\mpp{n}{m-2} \\
&+\frac{1}{2}n\,m\lp\Xi_{\phi\pi}+\Xi_{\pi\phi}\rp\mpp{n-1}{m-1}\,.
\end{split}
\ee
Note that in the third line, we have explicitly written the sum over the expansion of $(\ln V)_{,\phi}$ starting at 1 since the zeroth order was cancelled by the $D_\pi$ term in (\ref{eq:dmpp}). In analogy, we could expand the other contributions of the potential as
\be
\mppp{m}{n}{(\ln V)_{,\phi}}=\sum_{i=0}\frac{(\ln V)^{(i+1)}}{i!}\mpp{m+i}{n}\,,
\ee
where the sum begins at zeroth order.

For illustrative purpose, we have shown in Fig. \ref{fig:PDF} the comparison between a Gaussian and a generalized PDF
\be
\label{eq:generalPDF}
P(x)=\frac{1}{2^{\frac{3-k_4}{2-k_4}}\Gamma\lb\frac{3-k_4}{2-k_4}\rb\sigma}\lp1+\text{Erf}\lb s_3\frac{x}{\sqrt{2\sigma^2}}\rb\rp \exp\lb-\frac{1}{2}\left\vert \frac{x}{\sigma}\right\vert^{2-k_4}\rb\,,
\ee
where $s_3$ and $k_4$ parametrize the skewness and kurtosis respectively. Notice that for $s_3=k_4=0$ one recovers a normal distribution.
%-------------
%Stochastic Slow-roll inflation
%-------------
\section{Stochastic slow-roll inflation}
\label{app:StochasticSR}

Within slow-roll inflation, the stochastic dynamics of the inflaton are governed by a single Langevin equation
\be
\frac{d\phi}{dN}=D_\phi+\xi_\phi\,.
\ee
Using the Fokker-Planck equation we can compute the evolution for any $n$-point correlation
\begin{align}
\frac{d\mphi{n}}{dN}&=n\mpphi{n-1}{D_\phi}-n\mphi{n-1}D_\phi+\frac{1}{2}n(n-1)\mphi{n-2}\Xi_{\phi\phi} \\
&=n\sum_{i=0}^{p-n}\frac{D_\phi^{(i+1)}}{(i+1)!}\mphi{n+i}+\frac{1}{2}n(n-1)\mphi{n-2}\Xi_{\phi\phi}\,,
\end{align}
where $p$ is the maximum order considered. If $D_\phi$ is a polynomial, $p$ is determined by its order. Otherwise, we have to truncate the serie at a given order.

If we want to compute the 2-point correlation to its lower order, we have to solve
\be
\frac{d\mphi{2}}{dN}=2D'_{\phi}\mphi{2}+\Xi_{\phi\phi}\,.
\ee
Recalling that the classical solution is given by $d\phi/dN=D_\phi$, we have
\be
\frac{d\mphi{2}}{dN}-\frac{2}{D_\phi}\frac{dD_{\phi}}{dN}\mphi{2}=\Xi_{\phi\phi}\,,
\ee
whose solution is
\be
\mphi{2}=D_\phi^2\int\frac{\Xi_{\phi\phi}}{D_\phi^2}dN\,.
\ee
Using that $D_\phi^2=2\eps/\kappa^2$ one can show that
\be
\frac{d\mphi{2}}{dN}-\epss\mphi{2}=\Xi_{\phi\phi}\,,
\ee
and therefore 
\be
\frac{d\zeta^2}{dN}=\frac{d}{dN}\lp\frac{\mphi{2}}{\phi'^2}\rp=\frac{\kappa^2\Xi_{\phi\phi}}{2\eps}\,.
\ee
Recalling that $\Xi_{\phi\phi}=(1-\eps)\PS_{\delta\phi}$, then
\be
\PS_\zeta=\frac{d\langle\zeta(N)\rangle^2}{d\ln k}=\frac{1}{1-\eps}\frac{d}{dN}\lp\frac{\mphi{2}}{\phi'^2}\rp=\frac{\kappa^2\PS_{\delta\phi}}{2\eps}\,,
\ee
which is the standard result for the power spectrum.

We can also compute the noise analytically from the mode equation, which reads
\begin{align}
\frac{d^2\phi_k}{dN^2}+(3-\epsilon)\frac{d\phi_k}{dN}+\lp \frac{V''(\phi)}{H^2}+\lp\frac{k}{aH}\rp^2\rp\phi_k&=0\,.
\end{align}
At leading order in SR, $H=\mathrm{const}+\order{\epsilon}$, and for a free field, $V=m^2\phi^2/2$, there is an exact solution, in terms of the Hankel functions $H^{(1)}_\nu$,
\be
\phi_k=\frac{\sqrt{\pi}}{2a}\lp\frac{1}{aH}\rp^{1/2}H^{(1)}_\nu\lb\frac{k}{aH}\rb
\ee
which can be approximated after horizon crossing to
\be
\vert\phi_k\vert\rightarrow\frac{H}{\sqrt{2k^3}}\lp\frac{k}{aH}\rp^{\frac{3}{2}-\nu}
\ee
with $\nu^2=9/4-m^2/H^2$. Thus,
\be
\pi_k=\frac{d\phi_k}{dN}=\frac{H}{\sqrt{2k^3}}\lp\nu-\frac{3}{2}\rp\lp\frac{k}{aH}\rp^{\frac{3}{2}-\nu}+\order{\epsilon'}\simeq-\frac{H}{\sqrt{2k^3}}\lp\frac{m^2}{3H^2}\rp\lp\frac{k}{aH}\rp^{\frac{3}{2}-\nu}
\ee
using that $da/dN=a$. Note also that $m^2/3H^2=V''/\kappa^2V\equiv\eta_{_{SR}}$. This result shows that $\pi_k$ is higher order in SR. Therefore, to leading order in SR only $\Xi_{\phi\phi}$ will contribute to the noise by (recall $k_\sigma=\sigma a H$)
\be
\Xi_{\phi\phi}(N)=\frac{k_\sigma^3}{2\pi^2}\frac{d\ln k_\sigma}{dN}\frac{H^2}{2k_\sigma^3}\lp\frac{k_\sigma}{aH}\rp^{3-2\nu}=\frac{H^2}{4\pi^2}\sigma^{3-2\nu}\simeq\frac{H^2}{4\pi^2}\,,
\ee
where in the last equality we have used that at this order in SR $\sigma^{3-2\nu}=\sigma^{\order{\epsilon^2}}\simeq1$\,. In total, we recover the standard Langevin equation
\begin{align}
\frac{d\phi}{dN}&=-\frac{V_{,\phi}}{\kappa^2 V}+\frac{H}{2\pi}\,\xi\,,
\end{align}
where the noise is normalized as $\langle\xi(N)\xi(N')\rangle=\delta(N-N')$.

%STATISTICS
\section{Statistics of PBH formation}
\label{app:statistics}

Due to its relevance, we summarize the derivation of the exact formula for $P(\zeta>\zeta_c)$ presented in Eq. (\ref{eq:ProbabilityPBH}). We start the derivation from the last equality of (\ref{eq:ProbabilityAbove})
\be
P(\zeta>\zeta_c)=\int_{\zeta_c}^{\infty}da\int_{-\infty}^\infty\frac{du}{2\pi}e^{-iua}\chi(u)=\int_{\zeta_c}^{\infty}da\int_{-\infty}^\infty\frac{du}{2\pi}\exp\lb{\sum_{p=1}^{\infty}\frac{i^p}{p!}}\kappa_{p}u_\phi^{p}\rb e^{-iua}\,,
\ee
where we have rewritten the characteristic function in terms of the cumulants following (\ref{eq:SingleCharacteristicFunction}). 

The first trick of the derivation is to rewrite all the dependence in $u$ as Gaussian that we can later integrate. For that, one can use the simple identity
\be
ue^{-iua}=+i\frac{\partial}{\partial a}e^{-iua}
\ee
to arrive at
\be
P(\zeta>\zeta_c)=\int_{\zeta_c}^{\infty}da\int_{-\infty}^\infty\frac{du}{2\pi}\exp\lb{\sum_{p=3}^{\infty}\frac{(-1)^p}{p!}}\kappa_{p}\frac{\partial^p}{\partial a^p}\rb e^{-\frac{1}{2}\kappa_2 u^2-i(a-\kappa_1)u}\,,
\ee
which can be integrated (using the standard change of variables) in $u$
\be
P(\zeta>\zeta_c)=\frac{1}{\sqrt{2\pi\kappa_2}}\int_{\zeta_c}^{\infty}da\exp\lb{\sum_{p=3}^{\infty}\frac{(-1)^p}{p!}}\kappa_{p}\frac{\partial^p}{\partial a^p}\rb e^{-\frac{(a-\kappa_1)^2}{2\kappa_2}}\,.
\ee

The second trick of the derivation is to expand the exponential with the partial derivatives in terms of a unique series that separates the Gaussian and non-Gaussian contributions
\be
\label{eq:PnewCumulants}
\begin{split}
P(\zeta>\zeta_c)&=\frac{1}{\sqrt{2\pi}}\int_{z_c}^{\infty}dz\sum_{n=0}^{\infty}\frac{1}{n!}\lp\sum_{p=3}^{\infty}\frac{(-1)^p}{p!}\bar{\kappa}_{p}\frac{\partial^p}{\partial z^p}\rp^n e^{-\frac{z^2}{2}} \\
&=\frac{1}{\sqrt{2\pi}}\int_{z_c}^{\infty}dz\lp1+\sum_{n=3}^{\infty}\frac{(-1)^n}{n!}\bar{K}_{n}\frac{\partial^n}{\partial z^n}\rp e^{-\frac{z^2}{2}}\,.
\end{split}
\ee
Here we have first changed variables to $z=(a-\kappa_1)/\kappa_2^{1/2}$, $z_c=(\zeta_c-\kappa_1)/\kappa_2^{1/2}$ and normalized $\bar{\kappa}_n=\kappa_n/\kappa_2^{n/2}$. Then, we have expanded the serie and defined new functions $\bar{K}_n$ that accounts for the different combinations of normalized cumulants. Note that $\bar{K}_n$ only starts to differ from $\bar{\kappa}_n$ at sixth order, when the terms from $n=2$ of the first line of (\ref{eq:PnewCumulants}) appear. 
In this way, and recalling the definition of the complementary  error function 
\be
\text{Erfc}(x)=\frac{2}{\sqrt{\pi}}\int_x^{\infty}e^{-z^2}dz
\ee
and of the Hermite polynomials 
\be
H_n(x)\equiv(-1)^ne^{x^2}\frac{d^n}{dx^n}e^{-x^2}\,,
\ee
one can arrive to the final result (\ref{eq:ProbabilityPBH}). While the identification of the Erfc is direct from the first term of (\ref{eq:PnewCumulants}), the Hermite polynomials can be introduced by noting that
\be
\int_{z_c}^\infty\frac{\partial^n}{\partial z^n}e^{-\frac{z^2}{2}}=\lb\frac{\partial^{n-1}}{\partial z^{n-1}}e^{-\frac{z^2}{2}}\rb^{z=\infty}_{z=z_c}=\frac{e^{-\frac{z_c^2}{2}}}{(-1)^n 2^{(n-1)/2}}H_{n-1}\lb\frac{z_c}{\sqrt{2}}\rb\,,
\ee
since the contribution of the integral at infinity vanishes. With this, the derivation of (\ref{eq:ProbabilityPBH}) is completed.

%-----------------
%-----------------
%Bib
\bibliographystyle{JHEP.bst}
\bibliography{Bib_StochasticInflation}

\providecommand{\href}[2]{#2}\begingroup\raggedright\begin{thebibliography}{10}

\bibitem{Carr:1974nx}
B.~J. Carr and S.~W. Hawking, {\it {Black holes in the early Universe}},  {\em
  Mon. Not. Roy. Astron. Soc.} {\bf 168} (1974) 399--415.

\bibitem{Carr:2016drx}
B.~Carr, F.~Kuhnel, and M.~Sandstad, {\it {Primordial Black Holes as Dark
  Matter}},  {\em Phys. Rev.} {\bf D94} (2016), no.~8 083504,
  [\href{http://arxiv.org/abs/1607.06077}{{\tt arXiv:1607.06077}}].

\bibitem{Garcia-Bellido:2017fdg}
J.~Garc{\'\i}a-Bellido, {\it {Massive Primordial Black Holes as Dark Matter and
  their detection with Gravitational Waves}},  {\em J. Phys. Conf. Ser.} {\bf
  840} (2017), no.~1 012032, [\href{http://arxiv.org/abs/1702.08275}{{\tt
  arXiv:1702.08275}}].

\bibitem{Sasaki:2018dmp}
M.~Sasaki, T.~Suyama, T.~Tanaka, and S.~Yokoyama, {\it {Primordial black
  holes---perspectives in gravitational wave astronomy}},  {\em Class. Quant.
  Grav.} {\bf 35} (2018), no.~6 063001,
  [\href{http://arxiv.org/abs/1801.05235}{{\tt arXiv:1801.05235}}].

\bibitem{Bird:2016dcv}
S.~Bird, I.~Cholis, J.~B. Mu{\~n}oz, Y.~Ali-Ha{\"\i}moud, M.~Kamionkowski,
  E.~D. Kovetz, A.~Raccanelli, and A.~G. Riess, {\it {Did LIGO detect dark
  matter?}},  {\em Phys. Rev. Lett.} {\bf 116} (2016), no.~20 201301,
  [\href{http://arxiv.org/abs/1603.00464}{{\tt arXiv:1603.00464}}].

\bibitem{Clesse:2016vqa}
S.~Clesse and J.~Garc{\'\i}a-Bellido, {\it {The clustering of massive
  Primordial Black Holes as Dark Matter: measuring their mass distribution with
  Advanced LIGO}},  {\em Phys. Dark Univ.} {\bf 15} (2017) 142--147,
  [\href{http://arxiv.org/abs/1603.05234}{{\tt arXiv:1603.05234}}].

\bibitem{Sasaki:2016jop}
M.~Sasaki, T.~Suyama, T.~Tanaka, and S.~Yokoyama, {\it {Primordial Black Hole
  Scenario for the Gravitational-Wave Event GW150914}},  {\em Phys. Rev. Lett.}
  {\bf 117} (2016), no.~6 061101, [\href{http://arxiv.org/abs/1603.08338}{{\tt
  arXiv:1603.08338}}].

\bibitem{Clesse:2015wea}
S.~Clesse and J.~Garc{\'\i}a-Bellido, {\it {Massive Primordial Black Holes from
  Hybrid Inflation as Dark Matter and the seeds of Galaxies}},  {\em Phys.
  Rev.} {\bf D92} (2015), no.~2 023524,
  [\href{http://arxiv.org/abs/1501.07565}{{\tt arXiv:1501.07565}}].

\bibitem{Carr:2018rid}
B.~Carr and J.~Silk, {\it {Primordial Black Holes as Generators of Cosmic
  Structures}},  \href{http://arxiv.org/abs/1801.00672}{{\tt
  arXiv:1801.00672}}.

\bibitem{Clesse:2017bsw}
S.~Clesse and J.~Garc{\'\i}a-Bellido, {\it {Seven Hints for Primordial Black
  Hole Dark Matter}},  \href{http://arxiv.org/abs/1711.10458}{{\tt
  arXiv:1711.10458}}.

\bibitem{Kopp:2010sh}
M.~Kopp, S.~Hofmann, and J.~Weller, {\it {Separate Universes Do Not Constrain
  Primordial Black Hole Formation}},  {\em Phys. Rev.} {\bf D83} (2011) 124025,
  [\href{http://arxiv.org/abs/1012.4369}{{\tt arXiv:1012.4369}}].

\bibitem{Harada:2013epa}
T.~Harada, C.-M. Yoo, and K.~Kohri, {\it {Threshold of primordial black hole
  formation}},  {\em Phys. Rev.} {\bf D88} (2013), no.~8 084051,
  [\href{http://arxiv.org/abs/1309.4201}{{\tt arXiv:1309.4201}}]. [Erratum:
  Phys. Rev.D89,no.2,029903(2014)].

\bibitem{Young:2014ana}
S.~Young, C.~T. Byrnes, and M.~Sasaki, {\it {Calculating the mass fraction of
  primordial black holes}},  {\em JCAP} {\bf 1407} (2014) 045,
  [\href{http://arxiv.org/abs/1405.7023}{{\tt arXiv:1405.7023}}].

\bibitem{Germani:2018jgr}
C.~Germani and I.~Musco, {\it {The abundance of primordial black holes depends
  on the shape of the inflationary power spectrum}},
  \href{http://arxiv.org/abs/1805.04087}{{\tt arXiv:1805.04087}}.

\bibitem{Yoo:2018kvb}
C.-M. Yoo, T.~Harada, J.~Garriga, and K.~Kohri, {\it {PBH abundance from random
  Gaussian curvature perturbations and a local density threshold}},
  \href{http://arxiv.org/abs/1805.03946}{{\tt arXiv:1805.03946}}.

\bibitem{Garcia-Bellido:2017mdw}
J.~Garcia-Bellido and E.~Ruiz~Morales, {\it {Primordial black holes from single
  field models of inflation}},  {\em Phys. Dark Univ.} {\bf 18} (2017) 47--54,
  [\href{http://arxiv.org/abs/1702.03901}{{\tt arXiv:1702.03901}}].

\bibitem{Ezquiaga:2017fvi}
J.~M. Ezquiaga, J.~Garcia-Bellido, and E.~Ruiz~Morales, {\it {Primordial Black
  Hole production in Critical Higgs Inflation}},  {\em Phys. Lett.} {\bf B776}
  (2018) 345--349, [\href{http://arxiv.org/abs/1705.04861}{{\tt
  arXiv:1705.04861}}].

\bibitem{Kannike:2017bxn}
K.~Kannike, L.~Marzola, M.~Raidal, and H.~Veerm{\"a}e, {\it {Single Field
  Double Inflation and Primordial Black Holes}},  {\em JCAP} {\bf 1709} (2017),
  no.~09 020, [\href{http://arxiv.org/abs/1705.06225}{{\tt arXiv:1705.06225}}].

\bibitem{Ballesteros:2017fsr}
G.~Ballesteros and M.~Taoso, {\it {Primordial black hole dark matter from
  single field inflation}},  {\em Phys. Rev.} {\bf D97} (2018), no.~2 023501,
  [\href{http://arxiv.org/abs/1709.05565}{{\tt arXiv:1709.05565}}].

\bibitem{Cicoli:2018asa}
M.~Cicoli, V.~A. Diaz, and F.~G. Pedro, {\it {Primordial Black Holes from
  String Inflation}},  \href{http://arxiv.org/abs/1803.02837}{{\tt
  arXiv:1803.02837}}.

\bibitem{Ozsoy:2018flq}
O.~{\"O}zsoy, S.~Parameswaran, G.~Tasinato, and I.~Zavala, {\it {Mechanisms for
  Primordial Black Hole Production in String Theory}},
  \href{http://arxiv.org/abs/1803.07626}{{\tt arXiv:1803.07626}}.

\bibitem{Dalianis:2018frf}
I.~Dalianis, A.~Kehagias, and G.~Tringas, {\it {Primordial Black Holes from
  $\alpha$-attractors}},  \href{http://arxiv.org/abs/1805.09483}{{\tt
  arXiv:1805.09483}}.

\bibitem{Germani:2017bcs}
C.~Germani and T.~Prokopec, {\it {On primordial black holes from an inflection
  point}},  {\em Phys. Dark Univ.} {\bf 18} (2017) 6--10,
  [\href{http://arxiv.org/abs/1706.04226}{{\tt arXiv:1706.04226}}].

\bibitem{Motohashi:2017kbs}
H.~Motohashi and W.~Hu, {\it {Primordial Black Holes and Slow-Roll Violation}},
   {\em Phys. Rev.} {\bf D96} (2017), no.~6 063503,
  [\href{http://arxiv.org/abs/1706.06784}{{\tt arXiv:1706.06784}}].

\bibitem{Leach:2000yw}
S.~M. Leach and A.~R. Liddle, {\it {Inflationary perturbations near horizon
  crossing}},  {\em Phys. Rev.} {\bf D63} (2001) 043508,
  [\href{http://arxiv.org/abs/astro-ph/0010082}{{\tt astro-ph/0010082}}].

\bibitem{Leach:2001zf}
S.~M. Leach, M.~Sasaki, D.~Wands, and A.~R. Liddle, {\it {Enhancement of
  superhorizon scale inflationary curvature perturbations}},  {\em Phys. Rev.}
  {\bf D64} (2001) 023512, [\href{http://arxiv.org/abs/astro-ph/0101406}{{\tt
  astro-ph/0101406}}].

\bibitem{Starobinsky:1986fx}
A.~A. Starobinsky, {\it {STOCHASTIC DE SITTER (INFLATIONARY) STAGE IN THE EARLY
  UNIVERSE}},  {\em Lect. Notes Phys.} {\bf 246} (1986) 107--126.

\bibitem{Tolley:2008na}
A.~J. Tolley and M.~Wyman, {\it {Stochastic Inflation Revisited: Non-Slow Roll
  Statistics and DBI Inflation}},  {\em JCAP} {\bf 0804} (2008) 028,
  [\href{http://arxiv.org/abs/0801.1854}{{\tt arXiv:0801.1854}}].

\bibitem{Grain:2017dqa}
J.~Grain and V.~Vennin, {\it {Stochastic inflation in phase space: Is slow roll
  a stochastic attractor?}},  {\em JCAP} {\bf 1705} (2017), no.~05 045,
  [\href{http://arxiv.org/abs/1703.00447}{{\tt arXiv:1703.00447}}].

\bibitem{Vennin:2015hra}
V.~Vennin and A.~A. Starobinsky, {\it {Correlation Functions in Stochastic
  Inflation}},  {\em Eur. Phys. J.} {\bf C75} (2015) 413,
  [\href{http://arxiv.org/abs/1506.04732}{{\tt arXiv:1506.04732}}].

\bibitem{Assadullahi:2016gkk}
H.~Assadullahi, H.~Firouzjahi, M.~Noorbala, V.~Vennin, and D.~Wands, {\it
  {Multiple Fields in Stochastic Inflation}},  {\em JCAP} {\bf 1606} (2016),
  no.~06 043, [\href{http://arxiv.org/abs/1604.04502}{{\tt arXiv:1604.04502}}].

\bibitem{Vennin:2016wnk}
V.~Vennin, H.~Assadullahi, H.~Firouzjahi, M.~Noorbala, and D.~Wands, {\it
  {Critical Number of Fields in Stochastic Inflation}},  {\em Phys. Rev. Lett.}
  {\bf 118} (2017), no.~3 031301, [\href{http://arxiv.org/abs/1604.06017}{{\tt
  arXiv:1604.06017}}].

\bibitem{Saito:2008em}
R.~Saito, J.~Yokoyama, and R.~Nagata, {\it {Single-field inflation, anomalous
  enhancement of superhorizon fluctuations, and non-Gaussianity in primordial
  black hole formation}},  {\em JCAP} {\bf 0806} (2008) 024,
  [\href{http://arxiv.org/abs/0804.3470}{{\tt arXiv:0804.3470}}].

\bibitem{Byrnes:2012yx}
C.~T. Byrnes, E.~J. Copeland, and A.~M. Green, {\it {Primordial black holes as
  a tool for constraining non-Gaussianity}},  {\em Phys. Rev.} {\bf D86} (2012)
  043512, [\href{http://arxiv.org/abs/1206.4188}{{\tt arXiv:1206.4188}}].

\bibitem{Young:2013oia}
S.~Young and C.~T. Byrnes, {\it {Primordial black holes in non-Gaussian
  regimes}},  {\em JCAP} {\bf 1308} (2013) 052,
  [\href{http://arxiv.org/abs/1307.4995}{{\tt arXiv:1307.4995}}].

\bibitem{Young:2015cyn}
S.~Young, D.~Regan, and C.~T. Byrnes, {\it {Influence of large local and
  non-local bispectra on primordial black hole abundance}},  {\em JCAP} {\bf
  1602} (2016), no.~02 029, [\href{http://arxiv.org/abs/1512.07224}{{\tt
  arXiv:1512.07224}}].

\bibitem{Tada:2015noa}
Y.~Tada and S.~Yokoyama, {\it {Primordial black holes as biased tracers}},
  {\em Phys. Rev.} {\bf D91} (2015), no.~12 123534,
  [\href{http://arxiv.org/abs/1502.01124}{{\tt arXiv:1502.01124}}].

\bibitem{Young:2015kda}
S.~Young and C.~T. Byrnes, {\it {Signatures of non-gaussianity in the
  isocurvature modes of primordial black hole dark matter}},  {\em JCAP} {\bf
  1504} (2015), no.~04 034, [\href{http://arxiv.org/abs/1503.01505}{{\tt
  arXiv:1503.01505}}].

\bibitem{Franciolini:2018vbk}
G.~Franciolini, A.~Kehagias, S.~Matarrese, and A.~Riotto, {\it {Primordial
  Black Holes from Inflation and non-Gaussianity}},  {\em JCAP} {\bf 1803}
  (2018), no.~03 016, [\href{http://arxiv.org/abs/1801.09415}{{\tt
  arXiv:1801.09415}}].

\bibitem{Matarrese:1986et}
S.~Matarrese, F.~Lucchin, and S.~A. Bonometto, {\it {A Path Integral Approach
  To Large Scale Matter Distribution Originated By Nongaussian Fluctuations}},
  {\em Astrophys. J.} {\bf 310} (1986) L21--L26.

\bibitem{GarciaBellido:1996qt}
J.~Garcia-Bellido, A.~D. Linde, and D.~Wands, {\it {Density perturbations and
  black hole formation in hybrid inflation}},  {\em Phys. Rev.} {\bf D54}
  (1996) 6040--6058, [\href{http://arxiv.org/abs/astro-ph/9605094}{{\tt
  astro-ph/9605094}}].

\bibitem{Bullock:1996at}
J.~S. Bullock and J.~R. Primack, {\it {NonGaussian fluctuations and primordial
  black holes from inflation}},  {\em Phys. Rev.} {\bf D55} (1997) 7423--7439,
  [\href{http://arxiv.org/abs/astro-ph/9611106}{{\tt astro-ph/9611106}}].

\bibitem{Ivanov:1997ia}
P.~Ivanov, {\it {Nonlinear metric perturbations and production of primordial
  black holes}},  {\em Phys. Rev.} {\bf D57} (1998) 7145--7154,
  [\href{http://arxiv.org/abs/astro-ph/9708224}{{\tt astro-ph/9708224}}].

\bibitem{Yokoyama:1998pt}
J.~Yokoyama, {\it {Chaotic new inflation and formation of primordial black
  holes}},  {\em Phys. Rev.} {\bf D58} (1998) 083510,
  [\href{http://arxiv.org/abs/astro-ph/9802357}{{\tt astro-ph/9802357}}].

\bibitem{Pattison:2017mbe}
C.~Pattison, V.~Vennin, H.~Assadullahi, and D.~Wands, {\it {Quantum diffusion
  during inflation and primordial black holes}},  {\em JCAP} {\bf 1710} (2017),
  no.~10 046, [\href{http://arxiv.org/abs/1707.00537}{{\tt arXiv:1707.00537}}].

\bibitem{Biagetti:2018pjj}
M.~Biagetti, G.~Franciolini, A.~Kehagias, and A.~Riotto, {\it {Primordial Black
  Holes from Inflation and Quantum Diffusion}},
  \href{http://arxiv.org/abs/1804.07124}{{\tt arXiv:1804.07124}}.

\bibitem{Allahverdi:2006iq}
R.~Allahverdi, K.~Enqvist, J.~Garcia-Bellido, and A.~Mazumdar, {\it {Gauge
  invariant MSSM inflaton}},  {\em Phys. Rev. Lett.} {\bf 97} (2006) 191304,
  [\href{http://arxiv.org/abs/hep-ph/0605035}{{\tt hep-ph/0605035}}].

\bibitem{Namjoo:2012aa}
M.~H. Namjoo, H.~Firouzjahi, and M.~Sasaki, {\it {Violation of non-Gaussianity
  consistency relation in a single field inflationary model}},  {\em EPL} {\bf
  101} (2013), no.~3 39001, [\href{http://arxiv.org/abs/1210.3692}{{\tt
  arXiv:1210.3692}}].

\bibitem{Martin:2012pe}
J.~Martin, H.~Motohashi, and T.~Suyama, {\it {Ultra Slow-Roll Inflation and the
  non-Gaussianity Consistency Relation}},  {\em Phys. Rev.} {\bf D87} (2013),
  no.~2 023514, [\href{http://arxiv.org/abs/1211.0083}{{\tt arXiv:1211.0083}}].

\bibitem{Chen:2013aj}
X.~Chen, H.~Firouzjahi, M.~H. Namjoo, and M.~Sasaki, {\it {A Single Field
  Inflation Model with Large Local Non-Gaussianity}},  {\em EPL} {\bf 102}
  (2013), no.~5 59001, [\href{http://arxiv.org/abs/1301.5699}{{\tt
  arXiv:1301.5699}}].

\bibitem{Mukhanov:1985rz}
V.~F. Mukhanov, {\it {Gravitational Instability of the Universe Filled with a
  Scalar Field}},  {\em JETP Lett.} {\bf 41} (1985) 493--496. [Pisma Zh. Eksp.
  Teor. Fiz.41,402(1985)].

\bibitem{Sasaki:1986hm}
M.~Sasaki, {\it {Large Scale Quantum Fluctuations in the Inflationary
  Universe}},  {\em Prog. Theor. Phys.} {\bf 76} (1986) 1036.

\bibitem{Linde:1993xx}
A.~D. Linde, D.~A. Linde, and A.~Mezhlumian, {\it {From the Big Bang theory to
  the theory of a stationary universe}},  {\em Phys. Rev.} {\bf D49} (1994)
  1783--1826, [\href{http://arxiv.org/abs/gr-qc/9306035}{{\tt gr-qc/9306035}}].

\bibitem{Salopek:1990jq}
D.~S. Salopek and J.~R. Bond, {\it {Nonlinear evolution of long wavelength
  metric fluctuations in inflationary models}},  {\em Phys. Rev.} {\bf D42}
  (1990) 3936--3962.

\bibitem{GarciaBellido:1994vz}
J.~Garcia-Bellido, {\it {Jordan-Brans-Dicke stochastic inflation}},  {\em Nucl.
  Phys.} {\bf B423} (1994) 221--242,
  [\href{http://arxiv.org/abs/astro-ph/9401042}{{\tt astro-ph/9401042}}].

\bibitem{Winitzki:1999ve}
S.~Winitzki and A.~Vilenkin, {\it {Effective noise in stochastic description of
  inflation}},  {\em Phys. Rev.} {\bf D61} (2000) 084008,
  [\href{http://arxiv.org/abs/gr-qc/9911029}{{\tt gr-qc/9911029}}].

\end{thebibliography}\endgroup

\end{document}